\documentclass[12pt]{article}

\usepackage{scicite}

\usepackage{times}
\usepackage{graphicx}
\usepackage{amsmath}
\usepackage{braket}
\usepackage{siunitx}

\newenvironment{sciabstract}{%
\begin{quote} \bf}
{\end{quote}}

\title{A quantum-enhanced wide-field phase imager}

\author
{Robin Camphausen,$^{1\ast}$ \'Alvaro Cuevas,$^{1\dagger}$ Luc Duempelmann,$^{1}$ \\Roland A. Terborg,$^{1}$ Ewelina Wajs,$^{1}$ Simone Tisa,$^{2}$ \\Alessandro Ruggeri,$^{2}$ Iris Cusini,$^{3}$ Fabian Steinlechner,$^{4,5}$ \\Valerio Pruneri,$^{1,6\ddagger}$\\
\\
\normalsize{$^{1}$ICFO-Institut de Ciencies Fotoniques, The Barcelona Institute of Science and Technology,}\\
\normalsize{Av. Carl Friedrich Gauss, 3, 08860 Castelldefels, Barcelona, Spain}\\
\normalsize{$^{2}$Micro Photon Device SRL,}\\
\normalsize{Via Waltraud Gebert Deeg 3g, 39100, Bolzano, Italy}\\
\normalsize{$^{3}$Dipartimento di Elettronica, Informazione e Bioingegneria, Politecnico di Milano,}\\
\normalsize{Via Giuseppe Ponzio, 34, 20133 Milano, Italy}\\
\normalsize{$^{4}$Fraunhofer Institute for Applied Optics and Precision Engineering IOF,}\\
\normalsize{Albert-Einstein-Str. 7, 07745 Jena, Germany}\\
\normalsize{$^{5}$Abbe Center of Photonics, Friedrich Schiller University Jena,}\\
\normalsize{Albert-Einstein-Str. 6, 07745 Jena, Germany}\\
\normalsize{$^{6}$ICREA-Institució Catalana de Recerca i Estudis Avançats,}\\
\normalsize{Passeig Lluís Companys 23, 08010 Barcelona, Spain}
\\
\normalsize{$^\ast$Corresponding author. E-mail: robin.camphausen@icfo.eu,}\\
\normalsize{$^\dagger$E-mail: alvaro.cuevas@icfo.eu,}\\
\normalsize{$^\ddagger$E-mail: valerio.pruneri@icfo.eu.}
}

\date{}

\begin{document} 


\baselineskip24pt


\maketitle


\begin{sciabstract}
  Quantum techniques can be used to enhance the signal-to-noise ratio in optical imaging. Leveraging the latest advances in single photon avalanche diode array cameras and multi-photon detection techniques, here we introduce a super-sensitive phase imager, which uses space-polarization hyper-entanglement to operate over a large field-of-view without the need of scanning operation. We show quantum-enhanced imaging of birefringent and non-birefringent phase samples over large areas, with sensitivity improvements over equivalent classical measurements carried out with equal number of photons. The practical applicability is demonstrated by imaging a biomedical protein microarray sample. Our quantum-enhanced phase imaging technology is inherently scalable to high resolution images, and represents an essential step towards practical quantum imaging.
\end{sciabstract}


\section*{Introduction}

Entanglement can enhance precision measurements beyond the possibilities of classical optics \cite{giovannetti_quantum-enhanced_2004,pirandola_advances_2018}. This is of particular importance to applications that necessarily involve low photon flux, where shot noise becomes a limiting factor. Such a situation may be encountered in phase imaging of sensitive biological samples \cite{stephens_light_2003}, fragile quantum gas states \cite{eckert_quantum_2008} or atomic ensembles \cite{wolfgramm_entanglement-enhanced_2013}. In this case, it becomes attractive to perform phase imaging using N00N states consisting of N entangled photons between two optical modes, which are well known to yield a signal-to-noise ratio (SNR) enhancement of $\sqrt{N}$ over equivalent classical measurements \cite{giovannetti_quantum-enhanced_2004,pirandola_advances_2018,mitchell_super-resolving_2004,nagata_beating_2007,dowling_quantum_2008,giovannetti_advances_2011}. This effect is known as super-sensitivity.

Entanglement-enhanced phase imaging was demonstrated already for both birefringent \cite{israel_supersensitive_2014} and non-birefringent \cite{ono_entanglement-enhanced_2013} phase samples. However, neither of these works represent true imaging platforms as the entangled photons probing a sample were detected with single-pixel detectors and images were constructed by scanning the sample point-by-point. This inherently limits scalability as constructing a detailed image requires prohibitively long scanning times.

Here we show, for the first time, an entanglement-enabled super-sensitive phase imager operating in a wide-field configuration. By exploiting hyper-entanglement, that is simultaneous N00N state entanglement in the polarization degree of freedom and correlations in a massive pixel mode state space, our system is made scan-free. This enables the retrieval of phase information with a large field-of-view (FoV), using a single-photon sensitive SPAD array camera and computational methods adapted from digital holography. The holographic phase retrieval method employed has the advantage over conventional interferometric measurements that phases can be accurately retrieved regardless of sample absorbance and without requiring a priori knowledge of the illumination brightness and phase offset \cite{gabor_new_1948}. Our work is made possible through recent advances in quantum imaging \cite{moreau_imaging_2019,defienne_general_2018} and fabrication of single photon avalanche diode (SPAD) array cameras \cite{bronzi_100_2014}, allowing us to acquire spatially resolved multi-photon images with very high SNR. We demonstrate the experimental feasibility of our approach by retrieving precise phase images of birefringent and non-birefringent test samples, including a protein microarray sample which demonstrates the applicability for biomedical diagnostic applications. The ability to measure birefringent phase samples has also important applications in material science and crystallography \cite{sawyer_polymer_2008}. We show a sensitivity enhancement over equivalent classical measurements of $1.39\pm0.11$ and $1.25\pm0.06$, for the birefringent and non-birefringent samples respectively. Our method is inherently scalable to larger images with more pixels and represents an essential step towards a practically useful quantum-enhanced  biological and material inspection imaging platform.

\section*{Results}

\paragraph*{Quantum-enhanced large field-of-view phase imager}

Our system integrates a source of space-polarization hyper-entangled N00N states, a large FoV lens-free interferometric microscope (LIM) \cite{terborg_ultrasensitive_2016}, and a SPAD array camera, with computational methods for coincidence imaging and holographic phase retrieval. As can be seen in Fig.~1a, hyper-entangled photon pairs are generated by spontaneous parametric down conversion (SPDC) within a Sagnac interferometer (SI). Entanglement in the polarization degree of freedom is generated by combining the clockwise and counter-clockwise photon pair generations in the SI, resulting in a two-photon N00N state $(\ket{2_{H}0_{V}}+\ket{0_{H}2_{V}})/\sqrt{2}$, where $H$ and $V$ represent the horizontally and vertically polarized modes respectively \cite{steinlechner_efficient_2014}. The correlated nature of SPDC photon pair generation on the other hand yields space-momentum entanglement \cite{tasca_propagation_2009}. The near-field of the entangled state, where both photons are spatially correlated (approximately in the same spatial position) \cite{di_lorenzo_pires_near-field_2009}, is imaged onto a spatial light modulator (SLM), re-imaged into our LIM and detected by the SPAD array camera. After propagating through the entire setup, we express the quantum state as 
\begin{equation}
    \label{eq:NOON_afterLIM}
    \ket{\Psi} \approx \sum_{\mathbf{r} , \mathbf{r'}} \left[\ket{H}_{\mathbf{r}}\ket{H}_{\mathbf{r'}}+e^{i 2\Theta(\mathbf{r})}\ket{V}_{\mathbf{r}}\ket{V}_{\mathbf{r'}} \right],
\end{equation}
where we neglect normalisation coefficients for clarity. Here, $\mathbf{r}$ and $\mathbf{r'}$ are the transverse coordinates of the two spatially correlated photons, which are close in space and thus acquire approximately the same phase ($\Theta(\mathbf{r}) \approx \Theta(\mathbf{r'})$). The sample and setup therefore cause the two-photon entangled state to acquire a total phase difference between $H$ and $V$ of $2\Theta(\mathbf{r}) \approx \Theta(\mathbf{r}) + \Theta(\mathbf{r'})$. See Methods for details of the setup and Supplementary Material (SM) for the derivation of Eq.~(\ref{eq:NOON_afterLIM}). 

The LIM measures phase differences by interfering laterally displaced polarization states \cite{terborg_ultrasensitive_2016}, and can be used for the inspection of large area material and biological samples such as microarrays of proteins or micro-organisms for diagnostic applications \cite{yesilkoy_phase-sensitive_2018}. The crucial components of the LIM are two Savart plates (SPs; $SP_1$ and $SP_2$). For an input beam, $SP_1$ laterally displaces the $H$-polarized photons in one direction and the $V$-polarized photons in the orthogonal direction, thereby introducing a shear ($S$) between the two polarization components. Later, $SP_2$ is placed with an opposite orientation to $SP_1$ in order to revert this shear, which effectively forms a Mach-Zehnder interferometer (MZI) at each (lateral) spatial location, with the MZI modes separated from each other by the shear distance. Motorized tilting of $SP_1$ induces a controlled bias phase $\alpha$ between the two sheared spatial modes and associated polarization components after $SP_2$ \cite{terborg_ultrasensitive_2016,zhang_influences_2011}, over a large scanning range $0 < \alpha < 50 \pi$, with no measurable beam deviation. 

In the birefringent phase imaging configuration, the total phase $\Theta(\mathbf{r})$ after $SP_2$ equals to $\Theta_b(\mathbf{r}) = \phi_b (\mathbf{r}) + \alpha$, where $ \phi_b (\mathbf{r}) $ is a spatially dependent birefringent sample phase. This is illustrated in Fig.~1b, where the shown phase profile is always between $H$ and $V$ polarized light. On the other hand, for measuring a non-birefringent phase sample, the sample is placed between $SP_1$ and $SP_2$ of the LIM. In this configuration the LIM imprints a non-birefringent sample phase $ \phi_{nb} (\mathbf{r}) $ between the SPs onto a birefringent phase between $H$ and $V$ after $SP_2$. This results in the total phase $\Theta_{nb}(\mathbf{r}) = \phi_{nb} (\mathbf{r} + S/2) - \phi_{nb} (\mathbf{r} - S/2)  + \alpha$, where $S$ is the shear distance between $H$ and $V$ induced by the SPs, as  shown in Fig.~1c.

\paragraph*{Low-noise two-photon interference measurement using SPAD array camera}

The total phase factor $ 2 \Theta(\mathbf{r}) $ in Eq.~(\ref{eq:NOON_afterLIM}) acquired by the two-photon state $\ket{\Psi}$ is transformed into a measurable change in photon coincidences by projecting into the diagonal polarization bases. This is achieved by a half-wave plate (HWP) at $ \SI{22.5}{\degree }$ after $SP_2$, and a lateral displacement polarizing beam splitter (dPBS), which directs the photons with diagonal ($ D \equiv (H+V)/\sqrt{2}$) polarization to the left half of the camera sensor and those with anti-diagonal ($ A \equiv (H-V)/\sqrt{2}$) polarization to the right half. Using our SPAD array camera, we then measure spatially resolved photon coincidences \cite{defienne_general_2018} in the three possible polarization bases, corresponding to $\braket{DD|\Psi}$, $\braket{AA|\Psi}$, and $\braket{DA|\Psi}$ measurements (see Methods for details). We experimentally characterized, and then optimized the SNR of the measured coincidences, which depends on the entangled photon generation rate and camera acquisition parameters \cite{ndagano_imaging_2020, reichert_optimizing_2018} (see SM for details). We note that here a critical factor is the SPAD array camera's negligible read noise and very high frame rate, enabling the ultra low-noise acquisition of spatially resolved coincidences across 2048 spatial modes defined by the camera's pixels. The SPAD array camera is therefore an essential enabling technology for the super-sensitive phase imaging presented in this work.

\paragraph*{Entangled state characterization}

In order to characterize the quality of the entangled state we first acquired a series of coincidence count measurements (using Eq.~(\ref{eq:ccF}) from Methods section) while scanning through the LIM offset phase $\alpha$, with no sample present. In Fig.~2 we see that quantum two-photon interference manifests twice the periodicity of the single-photon interference, which is the expected signature of phase super-resolution for N00N state interference \cite{dowling_quantum_2008,mitchell_super-resolving_2004}. When integrating coincidences across the whole SPAD array camera (Fig.~2b), a relatively low fitted visibility of $\mathcal{V}_{overall}=0.670\pm0.022$ is obtained . However, when analysing the coincidences of one fixed pixel with those surrounding it (Fig.~2c), the fitted visibility is $\mathcal{V}_{local}=0.94\pm 0.06$. This result indicates high local fidelity with respect to the theoretical state $\ket{\Psi}$, validating the super-sensitive capabilities of our quantum resource \cite{israel_supersensitive_2014,slussarenko_unconditional_2017}. The discrepancy in visibility between Fig.~2b and c indicates a spatially-dependent phase background across the N00N state wave-front, which we characterize and remove when imaging samples (see SM for details).

\paragraph*{Holographic phase imaging from coincidence counts}

The processed coincidence counts $cc_F$, calculated from Eq.~(\ref{eq:ccF}) (in Methods), represent a four-dimensional quantity (two spatial dimensions per pixel). This was transformed into a two-dimensional coincidence image ($ci$) using the following mapping
\begin{equation} \label{eq:coincImage}
    ci(x,y) = \sum_{x'=1}^{W} \sum_{y'=1}^{Z} cc_F(x,y,x',y'),
\end{equation}
where $W$ ($Z$) is the image width (height), that is, the number of pixels in the $x$ ($y$) dimension of the camera sensor, and $cc_F(x,y,x',y')$ is the number of coincidences between pixels with coordinates $[x,y]$ and $[x',y']$. Eq.~(\ref{eq:coincImage}) produces a two-dimensional image with the same number of pixels as the camera sensor, representing the two-photon counts at each pixel (See SM for details).

Entanglement-enhanced phase images were retrieved using phase-shifting digital holography (PSDH) \cite{malacara_optical_2007}, taking advantage of the tunable LIM bias phase. Coincidence images were acquired for four different bias phases $\alpha = \{0, \pi/4, \pi/2, 3\pi/4\}$, where the spatially resolved coincidence image $ci(\mathbf{r}, \alpha)$ depends on the bias $\alpha$ and is calculated using Eq.~(\ref{eq:coincImage}). The sample phase image is then retrieved according to
\begin{equation}
\label{eq:PSDH_NOON}
    \hat{\phi}_{\mathrm{N00N}}(\mathbf{r}) = \frac{1}{2} \tan^{-1}\left[\frac{ci(\mathbf{r},\pi/4) - ci(\mathbf{r}, 3\pi/4)}{ci(\mathbf{r}, \pi/2) - ci(\mathbf{r}, 0)}\right],
\end{equation}
where the circumflex on $\hat{\phi}_{\mathrm{N00N}}$ indicates that it is an estimator of the sample phase, calculated from the experimental N00N state interference. In the birefringent phase imaging configuration, Eq.~(\ref{eq:PSDH_NOON}) estimates $\phi_b$, whereas in the non-birefringent phase imaging configuration, it estimates $\phi_{nb} (\mathbf{r} + S/2) - \phi_{nb} (\mathbf{r} - S/2)$, the sheared non-birefringent phase. The phase image is retrieved for each of the three polarization projections ($\bra{DD}$, $\bra{AA}$ and $\bra{DA}$) and then combined for a more accurate estimation.

\paragraph*{Super-sensitive imaging of birefringent sample}

We first investigated the entanglement-enhanced phase imaging capabilities of our system by measuring a birefringent test sample generated by the SLM (pattern shown in Fig.~3a). An equal number of photons was used to retrieve a phase image using classical (single-photon) intensity interference, and entanglement-enhanced (two-photon) N00N state interference. Over all four phase-shifted images a total of $I_{tot} = 818315$ single-photons and $ci_{tot} = 406428$ two-photon coincidences were detected, respectively (i.e. $I_{tot} \approx 2 ci_{tot}$) . Fig.~3b shows the classical phase estimate image $\hat{\phi}_{\mathrm{Classical}}$, calculated from Eq.~(\ref{eq:PSDH_intensity}) (in Methods). Fig.~3d on the other hand, shows the entanglement-enhanced phase estimate $\hat{\phi}_{\mathrm{N00N}}$, calculated from Eq.~(\ref{eq:PSDH_NOON}). Fig.~3c and e show cross-sections of the background noise in the classical and entanglement-enhanced phase estimate images, respectively. Fig.~3b and d both show the recovered sample phase well, whose accuracy is further confirmed with the zero-normalized cross-correlation image matching metric (details in SM). 

In order to quantify the sensitivity enhancement that our protocol provides, we compute the local uncertainty ($LU$) of the images, that is the root-mean-squared differences between all pairs of neighbouring pixels \cite{israel_supersensitive_2014}. The regions indicated by the black rectangles in Fig.~3b and d respectively were used to calculate the $LU$, yielding $LU_{\mathrm{Classical}} = 0.091 \pm 0.005$ and $LU_{\mathrm{N00N}} = 0.065 \pm 0.004$, where the errors in $LU$ represent the statistical standard error. We therefore obtained a reduction in noise from $\hat{\phi}_{\mathrm{Classical}}$ to $\hat{\phi}_{\mathrm{N00N}}$, which can be seen qualitatively by comparing the roughness of Fig.~3c and e, and numerically as $LU_{\mathrm{N00N}}/LU_{\mathrm{Classical}} = 0.72 \pm 0.06$. The above result is consistent with the predicted phase super-sensitivity for our system $ sd(\hat{\phi}_{\mathrm{N00N}})/sd(\hat{\phi}_{\mathrm{Classical}}) = 0.79 \pm 0.05$ (see Methods, or detailed calculation in SM), and close to the theoretical bound of $1/\sqrt{2}\approx 0.707$.

\paragraph*{Super-sensitive imaging of protein microarray sample}

A non-birefringent phase sample was implemented by fabricating a microarray of protein spots on a glass slide (details in Methods), similar to clinical microarray assays, where a range of capture antibodies are spotted onto a glass slide, each binding with a specific biomarker (e.g. an indicator of a disease). Measuring a change in signal for a given spot therefore confirms the presence or absence of a certain condition, aiding in rapid diagnosis \cite{stears_trends_2003}. Accurately imaging the phase jumps due to the presence or absence of proteins in such a biological sample, and showing a quantum enhancement in this measurement confirms the direct applicability of our entanglement-enhanced imaging system to diagnostics applications. As shown in Fig.~1c, the microarray test sample ($\phi_{nb}$) was inserted into the LIM for measuring. Fig.~4a shows a reference phase image retrieved under high intensity illumination, whereas in Fig.~4b and Fig.~4c are shown the low intensity illumination (single-photon) and entanglement-enhanced phase estimates $\hat{\phi}_{\mathrm{Classical}}$ and $\hat{\phi}_{\mathrm{N00N}}$, respectively. As with the birefringent sample, an equal number of photons detected was used to reconstruct the phase image estimates $\hat{\phi}_{\mathrm{Classical}}$ and $\hat{\phi}_{\mathrm{N00N}}$ ($I_{tot} = 3159383$ single-photons and $ci_{tot} = 1551916$ two-photon coincidences, i.e. $I_{tot} \approx 2 ci_{tot}$), permitting a fair comparison of phase sensitivity for the two methods. Horizontal cross-sections of the phase images (Fig.~4d-f) confirm the accuracy of the entanglement-enhanced measurement compared to both classical ones. The contrast between spots and surrounding background, indicating the presence and absence of proteins respectively, is clear in all measurements, which confirms the suitability of the technique for probing diagnostic microarrays. We again compare the $LU$, using the areas defined by the black rectangles. The extracted values are $LU_{\mathrm{Classical}} = 0.059 \pm 0.002$ and $LU_{\mathrm{N00N}} = 0.047 \pm 0.001$, which provides an enhancement of $LU_{\mathrm{N00N}}/LU_{\mathrm{Classical}} = 0.80 \pm 0.04$, again consistent with the predicted phase super-sensitivity of $0.79 \pm 0.05$ (see Methods, or detailed calculation in SM) and close to the theoretical bound of $1/\sqrt{2}\approx 0.707$.

\section*{Discussion}

It is straightforward to extend our protocol to larger N00N states with $N>2$ photons. Analogously to what is shown here, if all $N$ photons of the state are spatially correlated, a SPAD array camera can measure the $N>2$ multi-photon coincidence images, so the phase can be retrieved using PSDH with a theoretical sensitivity enhancement of $\sqrt{N}$, rather than the  $\sqrt{2}$ factor currently afforded by entangled photon pairs. However, using SPDC to generate N00N states with many photons is experimentally very challenging, with past demonstrations showing only probabilistic generation of N00N states with relatively small $N$ \cite{mitchell_super-resolving_2004,afek_high-noon_2010}. On the other hand, recent progress in quantum dot, and cavity based entangled photon sources has been promising \cite{raino_superfluorescence_2018,peniakov_towards_2020,maleki_high-n00n_2019}, and we are therefore hopeful that N00N state sources with high $N$ may become available in the future, which will enable greater sensitivity enhancements in our system. 

In our experiment, the image resolution of $32 \times 32$ pixels was only limited by the sensor of our SPAD array camera. Based on recent developments of SPAD arrays with up to megapixel resolution \cite{morimoto_megapixel_2020}, we expect that our protocol can be fully exploited for highly detailed quantum-enhanced phase imaging. We note that our protocol becomes susceptible to errors when the correlation width between photons is comparable to the sample feature size, as in this case it is likely that the two photons acquire different phases. This implies a breakdown in the approximations that lead to Eq.~(\ref{eq:NOON_afterLIM}) (details in SM), and leads to reduced edge contrast, as can be seen in the blurred edges of the ``$\phi$" pattern in Fig.~3d. Future work will focus on developing an entangled photon source with a tighter photon pair correlation width to improve spatial resolution.

Sensitivity comparisons between the classical and quantum-enhanced measurements were made here for an equal number of photons counted, with post-selection used for coincidence counting. Photon losses due to imperfect optical efficiencies were not taken into account, which is standard practice in almost all works on quantum-enhanced phase measurements to date \cite{mitchell_super-resolving_2004,nagata_beating_2007,israel_supersensitive_2014,ono_entanglement-enhanced_2013}. However, as N00N state phase measurements are highly sensitive to loss, a real sensitivity advantage can only be shown by comparing an equal number of photons at the sample \cite{slussarenko_unconditional_2017}, which necessitates very high optical efficiencies \cite{thomas-peter_real-world_2011}. To this end, future work will focus on developing entangled photon sources at shorter wavelengths in combination with enhanced efficiency SPAD array cameras \cite{sanzaro_single-photon_2018,gulinatti_custom_2021,ghioni_resonant-cavity-enhanced_2009}, in order to dramatically improve system optical efficiency. Another promising approach would be to use superconducting nanowire single-photon detectors (SNSPD), which routinely achieve detection efficiencies of close to unity \cite{reddy_superconducting_2020}. In particular, SNSPD image sensors have recently been demonstrated \cite{zhao_single-photon_2017,wollman_kilopixel_2019} which, while currently possessing some limitations, could in future be used for the required high efficiency coincidence image detection. Lastly, future work will also focus on modifying our method to achieve enhanced phase imaging using other quantum states that have less demanding optical efficiency requirements than N00N states \cite{higgins_entanglement-free_2007,thekkadath_quantum-enhanced_2020,kacprowicz_experimental_2010,braun_quantum-enhanced_2018}. We are therefore optimistic that the technological requirements for our method to yield a true quantum advantage over classical phase measurements will soon be met.

Note that despite also using holographic phase retrieval, the entanglement-enhanced microscope presented in this work is quite different from the quantum-enabled holography technique from Ref.~\cite{defienne_polarization_2021}. In that work two photons are spatially separated and non-local photon correlations are needed for holographic reconstruction. In our system on the other hand, both entangled photons pass through the sample together, which is the crucial aspect that enables our system to achieve super-sensitive phase imaging.

In conclusion, we have successfully implemented a practical large FoV, scan-free quantum-enhanced phase imaging protocol, capable of retrieving phase images with decreased noise compared to equivalent classical measurements.  Our system uses space-polarization hyper-entanglement, generated by an integrated source of quantum light, and combines a lens-free interferometric microscope with robust phase-scanning mechanism and novel data processing of images produced by a SPAD array camera. Polarization entanglement is exploited as a resource for phase super-sensitivity, while photon pair spatial correlations ensure that coincidence detections are confined to nearby pixels, thereby enabling scan-free simultaneous multi-photon imaging on many spatial modes across the whole FoV. For birefringent and non-birefringent phase samples we measured reductions in noise of the retrieved phase images, by factors of $0.72 \pm 0.06$ and $0.80 \pm 0.04$, whose inverse values yield the sensitivity enhancements of $ 1.39 \pm 0.11$ and $1.25 \pm 0.06$, respectively. Precise measurement of a protein micro-array demonstrate that bio-markers can be well identified. We expect systematic calibrations of the phase-response of specific samples to allow identifying bio-marker concentration. This advance shows compatibility of our quantum-enhanced method with medical diagnostic applications, with further use cases extending to a range of material and biological inspection tasks such as monitoring photoresist-based micro-fabrication, inspection of semiconductor and crystal materials, and observation of living organisms without inducing cellular damage or genetic degradation. We believe that with realistic future developments our technique will be highly competitive with respect to classical alternatives in which delicate samples cannot be analysed without risks of permanent damage under strong illumination, and that this work is thus an important step towards practically useful quantum imaging.

\section*{Materials and Methods} 

\paragraph*{Details of experimental setup}

As shown in Fig.~1a, a CW single-mode laser (Toptica TopMode) at $\SI{405.6}{nm}$ wavelength is used to pump a type-0 periodically poled Potassium Titanyl Phosphate (ppKTP) crystal inside the Sagnac interferometer. A superposition of SPDC processes takes places, which generates the quantum state $\ket{\Psi}$, composed of photon pairs at $\SI{811.2}{nm}$ wavelength \cite{steinlechner_efficient_2014}. The ppKTP crystal is temperature controlled using a Peltier oven in order to satisfy the degenerate phase matching condition. The laser power, measured before the SI, was fixed to \SI{3}{\milli\watt} for background measurements and fixed to \SI{0.6}{\milli\watt} for sample measurements. The entangled photon pairs are imaged onto the SLM (Holoeye Pluto-2) using two lenses of focal lengths $L_{1}=\SI{300}{\milli\meter}$ and $L_{2}=\SI{2500}{\milli\meter}$ in a 4f configuration, then re-imaged with two further lenses ($L_{3}=\SI{250}{\milli\meter}$ in the measurement of the birefringent sample, $L_{3}=\SI{500}{\milli\meter}$ in the measurement of the birefringent sample; $L_{4}=\SI{500}{\milli\meter}$) again in a 4f configuration into our LIM and SPAD camera (Micro Photon Devices SPC3). The LIM and camera are separated by less than the Rayleigh range of the imaging system. Therefore the SLM, the LIM and the SPAD camera are at conjugate planes of the SPDC plane, where photon pairs are spatially correlated \cite{di_lorenzo_pires_near-field_2009}. A $810\pm\SI{5}{\nano\meter}$ band-pass filter (BPF) is placed before the camera to remove environment noise and spurious pump light. Our SPAD camera is fitted with a microlens array, giving an effective pixel fill factor (FF) of $\approx75\%$. Therefore, the overall photon detection efficiency at $\SI{811.2}{nm}$ (taking into account FF) is approximately 3\%. The Savart plates (United Crystals) in our LIM induce a shear of \SI{450}{\micro\meter}.

\paragraph*{Coincidence counting using a camera}

Under low light conditions, where a pixel receives either zero or one photon, the coincidence counts $ \mathit{cc} $ between any two arbitrary pixels $ i $ and $ j $ can be calculated from $ N $ intensity image frames according to \cite{defienne_general_2018,ndagano_imaging_2020}
\begin{equation} \label{eq:cc}
    \mathit{cc}(i,j) = \sum_{l=1}^N I_{l,i} I_{l,j} - \frac{1}{N} \sum_{m,n=1}^N I_{m,i} I_{n,j},
\end{equation}
where $ I_{l,i} \in \{0,1\}$ represents the value returned by the $i^{th}$ pixel in the $l^{th}$ frame. The first term on the right side of Eq.~(\ref{eq:cc}) calculates the real and accidental coincidences across all frames, while the second term subtracts the accidentals, leaving only genuine photon coincidences \cite{defienne_general_2018}. In this work we calculate three cases for the above formula:
\begin{enumerate}
    \item $cc_{DD}(i,j)$ -- Pixels $i$ and $j$ both on the left half of the camera, i.e. a $\bra{DD}$ polarization projection,
    \item $cc_{AA}(i,j)$ -- Pixels $i$ and $j$ both on the right half of the camera, i.e. a $\bra{AA}$ polarization projection,
    \item $cc_{DA}(i,j)$ -- Pixels $i$ and $j$ on different halves of the camera, i.e. a $\bra{DA}$ polarization projection.
\end{enumerate}

In order to get accurate measurements of $\mathit{cc}(i,j)$, large numbers of frames ($N>10^6$) were used. The SPAD camera was operated at a frame rate of \SI{96}{\kilo\hertz}, with a deadtime of \SI{120}{\nano\second}. For background measurements we used an exposure time of \SI{70}{\nano\second} per frame, while for sample measurements we used an exposure time of \SI{10}{\nano\second} per frame. 

\paragraph*{SPAD array crosstalk removal}

Coincidence counts $cc_{DD}$ and $cc_{AA}$, as calculated by Eq.~(\ref{eq:cc}), also include crosstalk. In SPAD cameras, crosstalk coincidences $ \mathit{cc}_\mathrm{ct} (i,j) $ occur because after a real photon detection in pixel $i$, photons can be emitted from that location and detected by a nearby pixel $j$ with probability $ P_\mathrm{ct}(j|i) $  \cite{rech_optical_2008}, or vice versa \cite{eckmann_characterization_2020}. This is modelled by
\begin{eqnarray}
    \mathit{cc}_{\mathrm{ct}}(i,j)&=&\mathit{cc}_\mathrm{ct}(i|j)+\mathit{cc}_\mathrm{ct}(j|i)\nonumber\\
    &=&P_\mathrm{ct}(i|j)\sum_{l=1}^N I_{l,j} + P_\mathrm{ct}(j|i)\sum_{l=1}^N I_{l,i} \label{eq:xtalk}.
\end{eqnarray}
$\mathit{cc}_{DA}$ on the other hand, is not affected  by  crosstalk  because in this case the monitored pixels are far apart on  separate halves of the camera.

Characterizing $ P_{\mathrm{ct}}(i|j) $ is a highly non-trivial task, because it requires individual illumination of every single pixel. However, one can approximate the spatially dependent crosstalk to be uniform across the camera, depending only on the distance between two pixels \cite{lubin_quantum_2019,eckmann_characterization_2020}:
\begin{equation}\label{eq:xtalksimple}
    cc_{\mathrm{ct}}(i,j) = P_\mathrm{ct}(\Delta x, \Delta y) \sum_{l=1}^N (I_{l,j} + I_{l,i}),
\end{equation}
where $\Delta x = |x_i-x_j|$, $\Delta y=|y_i-y_j|$, with $ i $ and $j$ now explicitly expressed in terms of their $ x $ and $ y $ coordinates.

We characterized $P_\mathrm{ct}(\Delta x, \Delta y)$ by counting coincidences with the camera sensor covered, such that detections were only generated by dark counts. As dark counts of different pixels are uncorrelated, any coincidences measured are due to crosstalk \cite{ficorella_crosstalk_2016}. We then obtain the crosstalk probability:
\begin{equation}
    P_\mathrm{ct}(\Delta x, \Delta y) = \frac{1}{I_{tot}}\sum_{i=1}^{M}cc(i,i+\Delta\mathbf{r}),
\end{equation}
where $\Delta\mathbf{r} = [\Delta x, \Delta y]$, and $cc(i,i+\Delta\mathbf{r})$ is the coincidence count, calculated using Eq.~(\ref{eq:cc}) between the pixels with coordinates $[x_i,y_i]$ and $[x_i + \Delta x, y_i + \Delta y]$. $M$ is the total number of pixel pairs with equal separation $\Delta\mathbf{r}$, and $I_{tot}$ is the total number of detections on all pixels over all frames. See SM for an image of $P_\mathrm{ct}(\Delta x, \Delta y)$.

Therefore, the coincidences for the $\bra{DD}$ and $\bra{AA}$ polarization projections, with crosstalk subtracted, are calculated simply as $\mathit{cc}_{kk}(i,j)\equiv\mathit{cc}(i,j)-\mathit{cc}_{ct}(i,j)$, with $k=\{D,A\}$.

\paragraph*{Spatially uncorrelated noise removal}

When imaging coincidences additional sources of noise can appear due to the detection of spatially uncorrelated coincidences. However, these can be removed as with our experimental apparatus the probability of genuine coincidence detections becomes negligible for pixel pairs that are well-separated. We determine the SPDC photon pairs' spatial correlation width ($\sigma_{fit}$), which characterizes the maximum possible distance between correlated photons (see Ref.~\cite{defienne_quantum_2019}), by fitting the following Gaussian model to the measured coincidence counts:
\begin{equation}
\label{eq:gaussianFitAvg}
    G(i,j) = \exp \left[ \frac{-(((x_j-x_{0,i})^2 +(y_j-y_{0,i})^2)}{2 \sigma_{\mathrm{fit}}^2} \right],
\end{equation}
with  $[x_{0,i}, y_{0,i}] = [x_i, y_i]$ for $cc_{DD}$ and $cc_{AA}$. For $cc_{DA}$ we have $[x_{0,i}, y_{0,i}] = [x_i+d_x, y_i+d_y]$, where $d_x$ and $d_y$ are additional fitting parameters (details in SM). We then apply the following condition, yielding the filtered coincidence counts $cc_F$ with spatially uncorrelated noise coincidences to be removed:
\begin{equation}
\label{eq:ccF}
cc_F(i,j) =
    \begin{cases}
    cc(i,j) , &\text{if} \quad G(i,j) > t\\
    0, &\text{otherwise},
    \end{cases}
\end{equation}
where $cc(i,j)$ is calculated as described in the previous section for $cc_{DD}$, $cc_{AA}$ and $cc_{DA}$. $t$ is a threshold between 0 (no filtering) and 1 (filtering out all coincidences). For the phase imaging measurements in this work a threshold of $t = 0.5$ was used (details in SM).

\paragraph*{Phase-shifting interferometry method}

PSDH is a well-known technique for quantitative phase retrieval, which requires detecting four intensity images $I(\phi_{\mathrm{Sample}}, \alpha)$ with controlled offset phases ($\alpha = \{0, \pi/2, \pi, 3\pi/2\}$ for classical light) in order to access $\hat{\phi}_{\mathrm{Sample}}$ \cite{malacara_optical_2007}. This method originates from an algebraic inversion of the interference dependence function, and is usually written as
\begin{equation}
    \label{eq:PSDH_intensity}
    \hat{\phi}_{\mathrm{Sample}} = \tan^{-1}\left[\frac{I(\phi_{\mathrm{Sample}}, \pi/2) - I(\phi_{\mathrm{Sample}}, 3\pi/2)}{I(\phi_{\mathrm{Sample}}, \pi) - I(\phi_{\mathrm{Sample}}, 0)}\right].
\end{equation}
When measuring coincidence images as in our study, the N00N state quantum interference is governed by the phase $2\Theta = 2(\phi_{\mathrm{Sample}}+ \alpha)$ in Eq.~(\ref{eq:NOON_afterLIM}), where the factor of two represents the super-resolution enhancement. Accordingly, the four phase offsets are set to $\{0, \pi/4, \pi/2, 3\pi/4\}$, so that we can retrieve $2\hat{\phi}_{\mathrm{Sample}}$, which is therefore divided by two in order to get Eq.~(\ref{eq:PSDH_NOON}). To correctly set the four required offset phases, we first scanned through the interference curves by tilting $SP_2$ (shown in Fig.~2), which produces a continuous scan on $\alpha$. We then fitted a cosine to the data, to extract the right $SP_2$ tilts for the required offset phases.

\paragraph*{Modelling sensitivity enhancement}

To model the expected sensitivity enhancement we compare the noise (standard deviation) of the classical and entanglement-enhanced phase estimates $\hat{\phi}_{\mathrm{Classical}}$ and $\hat{\phi}_{\mathrm{N00N}}$, calculated using Eq.~\ref{eq:PSDH_intensity} and \ref{eq:PSDH_NOON} respectively. Using the error propagation formula \cite{ku_notes_1966}
\begin{equation}
    \label{eq:errorProp}
    \mathrm{sd}(f) = \sqrt{\sum_j \left[ \left(\frac{\partial f}{\partial x_j}\right)^2 \mathrm{sd}(x_j)^2 \right]},
\end{equation}
this noise can be related to the underlying measurements. For classical(N00N) interference the function $f$ corresponds to $\hat{\phi}_{\mathrm{Classical}}$($\hat{\phi}_{\mathrm{N00N}}$), and $x_j$ to the experimental intensity(coincidence) measurements. The standard deviations $\mathrm{sd}(x_j)$ of the intensity and coincidence image measurements, taking into account the real (i.e. imperfect) interference visibility and photon counting noise, can be expressed as
\begin{equation}\label{eq:sd_I}
    sd(I_j) = 
    \begin{cases}
    & \left[\frac{\mathcal{V}_{\mathrm{Classical}}\cos[\phi_{\mathrm{Sample}}(\mathbf{r}) + j\pi/2] + 1}{2} I_{\mathrm{tot}}/4 \right]^{1/2} \quad  \text{for $I_D$,}\\
    & \left[\frac{-\mathcal{V}_{\mathrm{Classical}}\cos[\phi_{\mathrm{Sample}}(\mathbf{r}) + j\pi/2] + 1}{2} I_{\mathrm{tot}}/4 \right]^{1/2} \quad  \text{for $I_A$,}
    \end{cases}
\end{equation}
where $I_D$($I_A$) is the intensity image measured in the diagonal(anti-diagonal) polarization basis (see SM for details), and 
\begin{equation}\label{eq:sd_ci}
    sd(ci_j) = 
    \begin{cases}
    & \kappa \big[\frac{\mathcal{V}\cos\left[2\phi_{\mathrm{Sample}}(\mathbf{r}) + j\pi/4\right] + 1}{4} ci_{\mathrm{tot}}/4 \big]^{1/2} \quad  \text{for $cc_{DD}$ and $cc_{AA}$,}\\
    & \kappa \big[\frac{-\mathcal{V}\cos\left[2\phi_{\mathrm{Sample}}(\mathbf{r}) + j\pi/4\right] + 1}{2} ci_{\mathrm{tot}}/4 \big]^{1/2} \quad  \text{for $cc_{DA}$.}
    \end{cases}
\end{equation}
Here $I_j$($ci_j$) is the intensity(coincidence) measurement corresponding to the $j^{th}$ step in the PSDH protocol, and $I_{tot}$($ci_{tot}$) is the total number of photons(coincidences) over all four steps. $\mathcal{V}_{\mathrm{Classical}}$($\mathcal{V}$) is the interference visibility for classical(N00N) interference, where we set $\mathcal{V}_{\mathrm{Classical}}$ to unity, while we measured $\mathcal{V} = 0.94\pm 0.06$. Some photon counting noise is added by the coincidence counting method used here \cite{ndagano_imaging_2020,reichert_optimizing_2018}, which is quantified by $\kappa \geq 1$ (where $\kappa=1$ for ideal shot noise limited measurements). We experimentally measure a value of $\kappa = 1.05$ (see SM for details).

For a fair sensitivity comparison we set the total number of photons used in the classical and the entanglement-enhanced phase estimation to be equal, i.e. $I_{tot} = 2 ci_{tot}$. Substituting Eq.~\ref{eq:sd_I} and \ref{eq:sd_ci} into Eq.~\ref{eq:errorProp} gives the standard deviations of the phase estimations retrieved for each polarization measurement, for both classical and quantum-enhanced methods. The phase retrievals due to the individual polarization measurements are then combined according to
\begin{align}
    \hat{\phi}_{\mathrm{Classical}} & = \frac{\hat{\phi}_{\mathrm{Classical, D}} + \hat{\phi}_{\mathrm{Classical, A}}}{2} \label{eq:combinephi_classical}\\
    \hat{\phi}_{\mathrm{N00N}} & = \frac{\hat{\phi}_{\mathrm{N00N, DD}} + \hat{\phi}_{\mathrm{N00N, AA}} + 2\hat{\phi}_{\mathrm{N00N, DA}}}{4}, \label{eq:combinephi_NOON}
\end{align}
which allows using the error propagation formula Eq.~\ref{eq:errorProp} again to obtain expressions for both $sd(\hat{\phi}_{\mathrm{Classical}})$ and $sd(\hat{\phi}_{\mathrm{N00N}})$ in terms of $I_{tot}$. Finally we numerically evaluate: $sd(\hat{\phi}_{\mathrm{N00N}})/sd(\hat{\phi}_{\mathrm{Classical}}) = 0.79 \pm 0.05$ (see SM for details).

\paragraph*{Protein microarray sample fabrication}

The microarray test sample was fabricated using commercially available Pierce Recombinant Protein A/G (Thermo Scientific 21186). First the stock solution at \SI{5}{\milli \gram}/mL was diluted using milli-Q water to a final concentration of \SI{500}{\micro \gram}/mL. This was then spotted (using a SCIENION sciFLEARRAYER S3 spotter) onto a borosilicate glass slide (NEXTERION Slide E, SCHOTT), coated with a multi-purpose epoxysilane layer that covalently binds most types of bio-molecules including amino- and non-modified DNA, RNA, and proteins. Spots of diameter \SI{500}{\micro \meter} were made with \SI{1000}{\micro \meter} centre-to-centre spacing, and the sample was left to dry overnight (24 hours) before measuring.


\nocite{procopio_geometry_2015,osorio_spatiotemporal_2008,tasca_propagation_2009,edgar_imaging_2012,terborg_ultrasensitive_2016,reichert_optimizing_2018,ku_notes_1966,avendano-alejo_optical_2006,di_stefano_zncc-based_2005}


\bibliographystyle{Science_modified}

\section*{Acknowledgements}
We thank Alexander Demuth for help with the design of components for the entangled photon source and for helpful discussions. We thank Sebastian Ecker for help with  setting up the entangled photon source.\\
\textbf{Funding:} This work has received funding from the European Union’s Horizon 2020 FET-Open research and innovation programme under grant agreement No. 801060 (Q-MIC project), from the European Union’s Horizon 2020 research and innovation programme under the Marie Skłodowska-Curie grant agreement No 713729 (ICFOstepstone 2), and from the Beatriu de Pinos-3 Postdoctoral Programme (BP3) under grant agreement ID 801370. This project has received funding from the European Union’s Horizon 2020 research and innovation programme under the Marie Skłodowska-Curie grant agreement No. 754510 (PROBIST). We acknowledge financial support from the Spanish State Research Agency through the ``Severo Ochoa” program for Centers of Excellence in R\&D (CEX2019-000910-S), and project TUNA-SURF (Grant No. PID2019-106892RB-I00), from Fundació Cellex, Fundació Mir-Puig, and from Generalitat de Catalunya through the CERCA program. \'Alvaro Cuevas' work was partially supported by Becas Chile N$^{\circ}$74200052 from the Chilean National Agency for Research and Development (ANID).\\
\textbf{Author Contributions:} V.P. proposed and directed the project. R.C. and A.C. built the setup and conducted the experiments. L.D., R.T. and R.C. developed essential hardware and software components. R.C. and A.C. developed the theory. E.W. fabricated the samples measured. S.T. and A.R. designed the microlens array for the SPAD camera and provided experimental support with the SPAD camera. I.C. contributed to further development of the SPAD array camera sensor. F.S. designed the entangled photon source. R.C. and A.C. wrote the paper with contributions from all authors.\\
\textbf{Competing interests:} V.P. is a co-assignee of U.S. Patent 8,472,031 B2, protecting part of the technology used in the experiments.\\
\textbf{Data and materials availability:} Essential data needed to evaluate the conclusions in this paper are present in the main manuscript and Supplementary Materials. Additional data related to this paper may be requested from the authors.



\clearpage

\begin{figure}[!htb]
	\centering
    \includegraphics[width=0.8\textwidth]{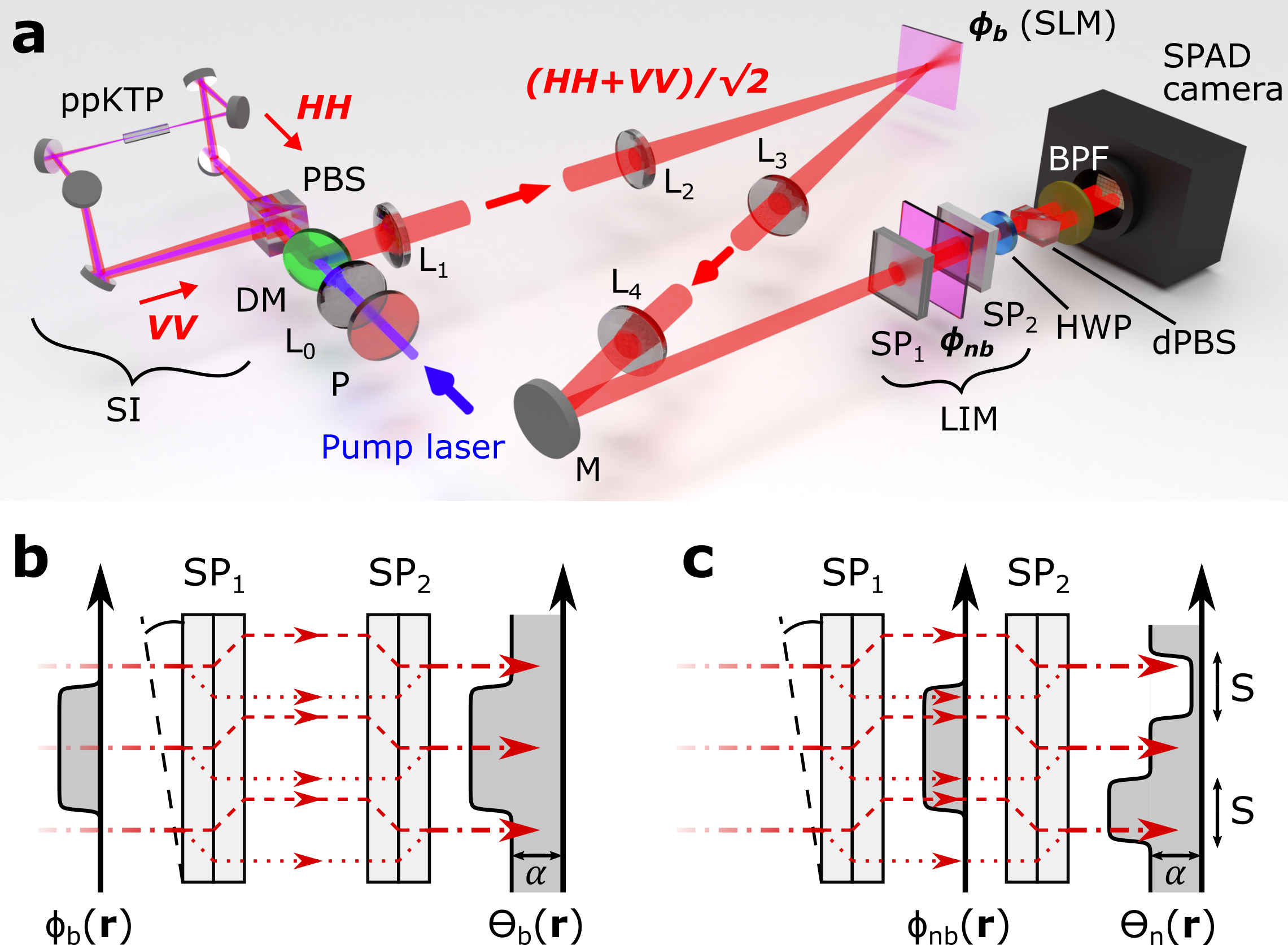}
	\caption{\textbf{Description of the experiment. (a)} Scheme of the entanglement-enhanced imaging setup. SI - Sagnac Interferometer, PBS - polarizing beam splitter, HWP - half-wave plate, L - lenses, DM - dichroic mirror, M - mirror, $\phi_b$ - birefringent sample (SLM), $\phi_{nb}$ - non-birefringent sample, SP - Savart plate, dPBS - lateral displacement polarizing beam splitter, BPF - band-pass filter. \textbf{(b)} Detecting birefringent phase samples with the LIM. \textbf{(c)} Detecting non-birefringent phase samples with the LIM. In \textbf{(b)} and \textbf{(c)} three example trajectories are shown through the LIM, dashed lines correspond to $H$, and dotted lines to $V$ polarized light. } 
\end{figure}

\begin{figure}[!htb]
	\centering
	\includegraphics[width=0.7\textwidth]{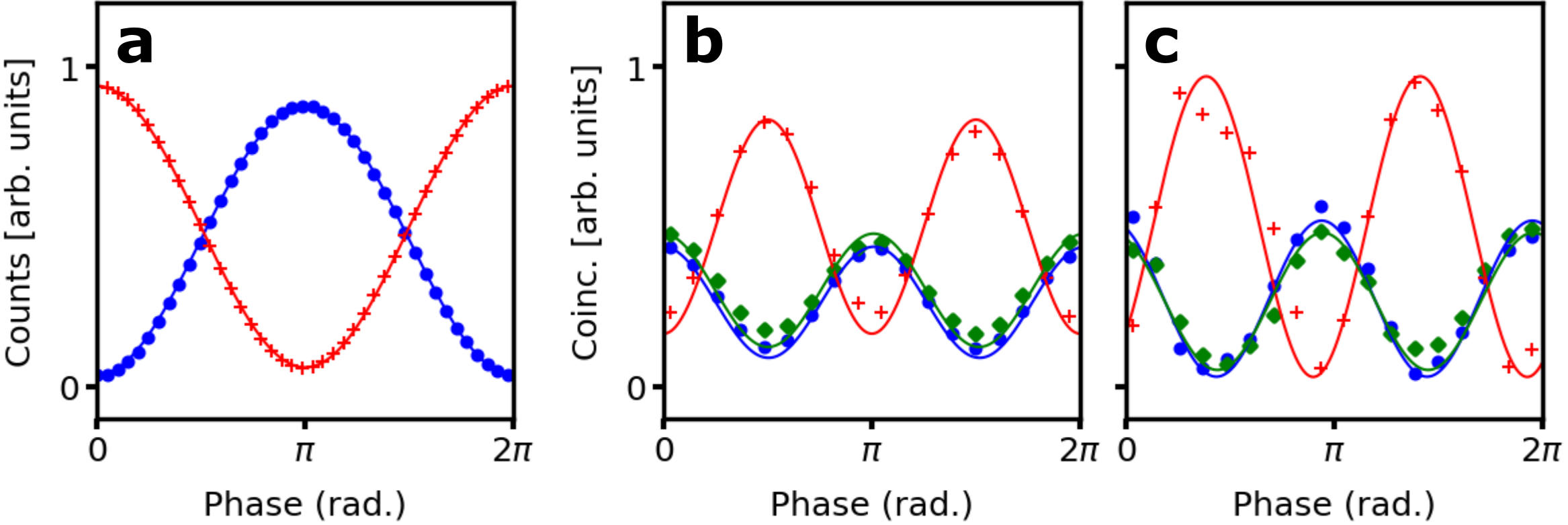}
	\caption{\textbf{Classical vs N00N state interference. (a)} Classical interference integrating across whole camera. Red crosses and blue circles correspond to $\bra{D}$ and $\bra{A}$ projections, respectively. \textbf{(b)} N00N state interference integrating across whole camera. \textbf{(c)} N00N state interference with a single fixed pixel. For \textbf{(b)} and \textbf{(c)}, red crosses, blue circles and green diamonds correspond to $\bra{DA}$, $\bra{DD}$ and $\bra{AA}$ projections, respectively. Solid lines are fitted curves.}
\end{figure}

\begin{figure*}
	\centering
	\includegraphics[width=\textwidth]{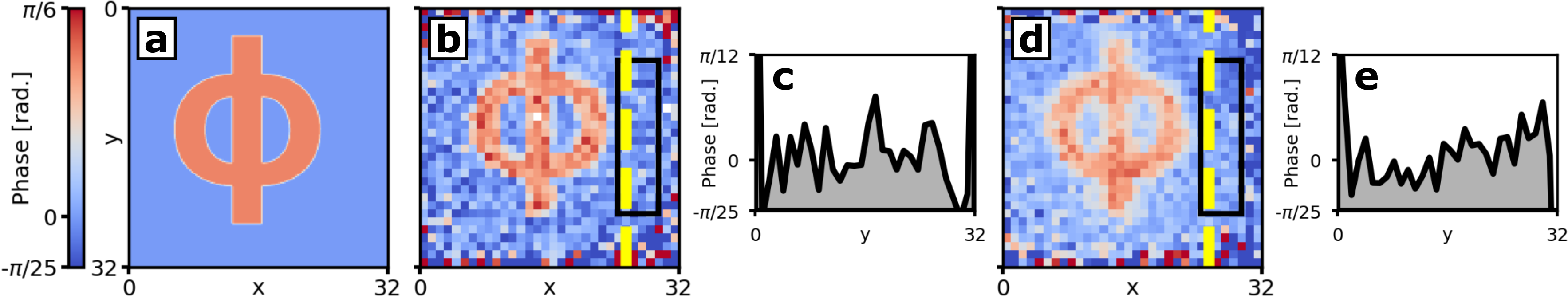}
	\caption{\textbf{Retrieved phase images of a birefringent sample. (a)} Phase profile applied to SLM. \textbf{(b)} Classical phase image $\hat{\phi}_{\mathrm{Classical}}$. \textbf{(c)} Cross-section of phase profile along yellow dashed line in \textbf{(b)}. \textbf{(d)} Entanglement-enhanced phase image $\hat{\phi}_{\mathrm{N00N}}$. \textbf{(e)} Cross-section of phase profile along yellow dashed line in \textbf{(d)}. Black rectangles in \textbf{(b)} and \textbf{(d)} indicate area used for $LU$ calculations. Clearly the pixel-to-pixel noise is reduced in \textbf{(d)} and \textbf{(e)} compared to \textbf{(b)} and \textbf{(c)}. The reduced edge contrast in \textbf{(d)} is due to the relatively large photon spatial correlation width, but can be addressed by engineering an entangled photon source with tighter spatial correlation.}
\end{figure*}

\begin{figure*}[htb]
	\centering
	\includegraphics[width=0.7\textwidth]{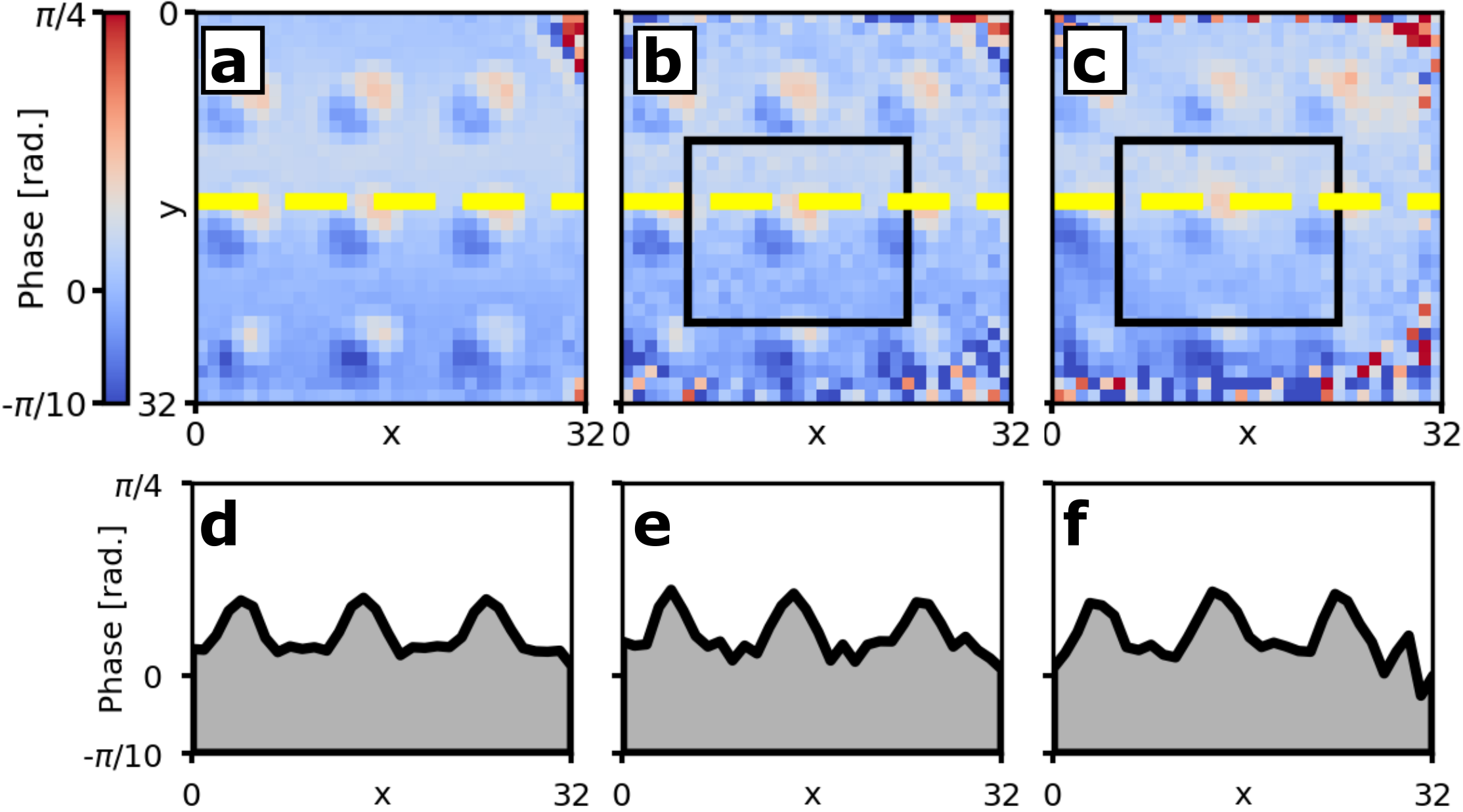}
	\caption{\textbf{Retrieved phase images of a non-birefringent protein microarray sample. (a)} Reference phase image from high intensity classical illumination. \textbf{(b)} Low intensity (single-photon level) classical illumination phase image $\hat{\phi}_{\mathrm{Classical}}$. \textbf{(c)} Entanglement-enhanced phase image $\hat{\phi}_{\mathrm{N00N}}$. \textbf{(d)}-\textbf{(f)} Cross-sections of phase profiles along yellow dashed lines in \textbf{(a)}-\textbf{(c)}. Black rectangles in \textbf{(b)} and \textbf{(c)} indicate area used for $LU$ calculations. All three experimental conditions show clear contrast between regions of high protein binding (circular spots) to regions with no binding (background). The entanglement-enhanced method \textbf{(c)} manifests less pixel-to-pixel noise than its classical counterpart \textbf{(b)}, for an equal number of photons detected.}
\end{figure*}

\clearpage

\section*{Supplementary materials}

\renewcommand{\theequation}{S\arabic{equation}}
\renewcommand{\thefigure}{S\arabic{figure}}
\renewcommand{\thesection}{S\arabic{section}}

\section{Derivation of two-photon wave function} \label{sect:derivEq1}

When looking at the $s$-signal and $i$-idler modes of a SPDC process, the generated state can be written as ({\it 51\/})
\begin{equation}
\ket{\psi_{SPDC}}=\ket{0_{s},0_{i}}+\chi\int d^{3}k_{s}\int d^{3}k_{i}\gamma(\omega_{s},\mathbf{k}_{s},\omega_{i},\mathbf{k}_{i})\mathbf{a}_{s}^{\dagger}(\mathbf{k}_{s})\mathbf{a}_{s}^{\dagger}(\mathbf{k}_{s})\ket{0_{s},0_{i}}
\end{equation}
where $\chi$ carries the interaction strength that creates the $s$ and $i$ photons. $\mathbf{k}_{n}$, $\omega_{n}$ and $\mathbf{a}_{n}^{\dagger}$ are the wave vector, frequency and creation operator for the photon in mode $n=[s,i]$, respectively. The relevant information concerning their spatial and spectral correlations is encoded in the $\gamma$-function, whose shape can be analyzed separately for the transverse and longitudinal components of the momentum, $\mathbf{k}=\mathbf{q}+k_{z}\mathbf{e}_{z}$. If the pump and SPDC beams are polarized and quasi-monochromatic ({\it 52\/}), for transverse photon detection only, the accessible correlation turns into

\begin{equation}
\label{eq:gamma_momentum}
    \gamma(\omega_{s},\mathbf{k}_{s},\omega_{i},\mathbf{k}_{i})\longrightarrow\gamma(\mathbf{q}_{s},\mathbf{q}_{i})= f(\mathbf{q}_{s}+\mathbf{q}_{i}) g(\mathbf{q}_{s}-\mathbf{q}_{i}),
\end{equation}
where $f$ and $g$ are non-identical Gaussian functions. The first term in Eq.~(\ref{eq:gamma_momentum}) is given by the spatial profile of the pump, while the second term is the Fourier transform of the phase-matching function ({\it 19\/}). Alternatively, the $\gamma$-function can also be expressed in terms of spatial coordinates by using the Fourier transformation 
\begin{equation}
\label{eq:gamma_position}
    \gamma(\mathbf{r}_{s},\mathbf{r}_{i})=\mathcal{N}\int d^{2}r_{s}\int d^{2}r_{i}\gamma(\mathbf{q}_{s},\mathbf{q}_{i})e^{i\mathbf{q}_{s}\cdot\mathbf{r}_{s}}e^{i\mathbf{q}_{i}\cdot\mathbf{r}_{i}},
\end{equation}
with $\mathcal{N}$ as a normalization factor. 

Now, if the photons are indistinguishable in all degrees of freedom, it is useful to re-write Eq.~(\ref{eq:gamma_momentum}) and Eq.~(\ref{eq:gamma_position}) without labels on the particles but only on their possible momenta and locations. For the momentum anti-correlation, one images the SPDC far-field plane into the SPAD array camera, which measures photon coincidences determined by ({\it 19, 53\/})
\begin{equation}
\label{eq:SPDCmomentum}
    \gamma(\mathbf{q}, \mathbf{q}') = \frac{\sigma_+ \sigma_-}{\pi}  \exp\left[- \frac{\sigma^2_+}{4}|\mathbf{q} + \mathbf{q}'|^2 - \frac{\sigma^2_-}{4}|\mathbf{q} - \mathbf{q}'|^2\right],
\end{equation}
where $\mathbf{q}$ and $\mathbf{q}'$ are the momenta of both photons, $\sigma_-$ and $\sigma_+^{-1} $ are the standard deviations of Gaussian functions describing the photon position and momentum correlations respectively, while $\sigma_+$ is the Gaussian waist of the pump laser spot ({\it 53\/}).

For the spatial correlation, one uses a 4f lens configuration to replicate and image the SPDC near-field (inside the non-linear crystal) into the SPAD array camera, which measures photon coincidences determined by ({\it 19, 53\/})
\begin{equation}
\label{eq:SPDC_space}
    \gamma(\mathbf{r}, \mathbf{r}') \equiv \frac{1}{\pi \sigma_+ \sigma_-}  \exp\left[- \frac{|\mathbf{r} + \mathbf{r}'|^2}{4\sigma^2_+} - \frac{|\mathbf{r} - \mathbf{r}'|^2}{4\sigma^2_-}\right],
\end{equation}

where $\mathbf{r}$ and $\mathbf{r}'$ are the spatial coordinates that each photon can take. The irradiance, that is the single photon intensity profile, is obtained simply by tracing out one of the two photons in Eq.~(\ref{eq:SPDC_space}). Imaging the SPDC emission near-field using a 4f lens configuration replicates this wave function at the image plane. As this is the case in our setup, Eq.~(\ref{eq:SPDC_space}) is therefore an accurate representation of the spatial two-photon distribution (neglecting magnification for simplicity) at the SLM, and at the LIM and camera plane. Moreover, unlike classical light, SPDC processes generated by a large pump spot in a bulk nonlinear crystal satisfy the condition $\sigma_-/\sigma_+ \ll 1$ ({\it 19\/}).

Our entanglement source produces two SPDC generations within the SI: one creates $V$ polarized photons in the counter-clockwise direction, and the other one creates $H$ polarized photons in the clockwise direction. Both generations are coherently superposed and can be represented by the common state
\begin{equation}
\label{eq:NOON_nophase}
\ket{\psi}=\int d^{2}r \int d^{2}r'\left(\gamma_{HH}(\mathbf{r},\mathbf{r}')\mathbf{h}^{\dagger}(\mathbf{r})\mathbf{h}^{\dagger}(\mathbf{r}')+\gamma_{VV}(\mathbf{r},\mathbf{r}')\mathbf{v}^{\dagger}(\mathbf{r})\mathbf{v}^{\dagger}(\mathbf{r}')\right)\ket{0}.
\end{equation}

Here, $\mathbf{h}^{\dagger}(\mathbf{r}_{n})$ and $\mathbf{v}^{\dagger}(\mathbf{r}_{n})$ are the photon creation operators for the horizontal and vertical polarization at spatial location $\mathbf{r}_{n}$. $\gamma_{HH} $ ($ \gamma_{VV} $) represents the spatial distribution according to Eq.~(\ref{eq:SPDC_space}) of the $HH$ ($VV$) generation with respective spatial correlation width $ \sigma_{HH,-}$ ($ \sigma_{VV,-}$) and momentum correlation $\sigma_{HH,+}^{-1} $ ($\sigma_{VV,+}^{-1}$). In general it is not the case that $ \sigma_{HH,-} = \sigma_{VV,-}$ and $\sigma_{HH,+}^{-1} = \sigma_{VV,+}^{-1}$. However, through careful alignment we assure that the $HH$ and $VV$ generations are indistinguishable not only in spatial profile, but also in momentum and position correlation (shown here in Section \ref{sect:alignment}), then allowing us to consider $\gamma_{HH}(\mathbf{r},\mathbf{r}')=\gamma_{VV}(\mathbf{r},\mathbf{r}')=\gamma(\mathbf{r},\mathbf{r}')$. Thus, after the state described in Eq.~(\ref{eq:NOON_nophase}) passes through a spatially dependent birefringent phase, as in general the two photons are not at the same spatial location, each one of them accumulates a different phase and the state can be written as
\begin{equation} 
\label{eq:NOONsimpler_phase}
    \ket{\psi} = \int d^{2}r \int d^{2}r'\gamma(\mathbf{r}, \mathbf{r}') \left( \mathbf{h}^{\dagger}(\mathbf{r})\mathbf{h}^{\dagger}(\mathbf{r}') + e^{i(\theta(\mathbf{r})+ \theta(\mathbf{r}'))}\mathbf{v}^{\dagger}(\mathbf{r})\mathbf{v}^{\dagger}(\mathbf{r}')\right)\ket{0},
\end{equation}

where $ \theta(\mathbf{r}) $ and $ \theta(\mathbf{r}') $ are the phases acquired by both photons at locations $\mathbf{r}$ and $\mathbf{r}'$. 

We note that in the SPDC near-field, the $\gamma$-function of Eq.~(\ref{eq:SPDC_space}) rapidly approaches zero when $ |\mathbf{r} - \mathbf{r}'| > \sigma_-$, that is, the two photons are most likely to be detected close to each other within a radius given by the position correlation width. If the spatially dependent phase varies on a length scale larger than the position correlation width, i.e. $ \theta(\mathbf{r} + \sigma_-) \approx \theta(\mathbf{r})$, both photons will acquire approximately the same phase $\theta(\mathbf{r})\approx\theta(\mathbf{r}')$, no matter their $\mathbf{r}$ and $\mathbf{r}'$ locations. Through careful optimization of our setup, we ensure that this condition is always fulfilled, so Eq.~(\ref{eq:NOONsimpler_phase}) can be reduced to the following polarization entangled N00N state in the image plane:
\begin{equation} 
\label{eq:NOONsimplest_phase}
    \ket{\psi} \approx \sum_{\mathbf{r},\mathbf{r}'} \left(\ket{H}_{\mathbf{r}}\ket{H}_{\mathbf{r}'} + e^{i2\Theta(\mathbf{r})}\ket{V}_{\mathbf{r}}\ket{V}_{\mathbf{r}'} \right),
\end{equation}

where we transform the integral into a sum due to discretization of modes in our detection system.

The factor of two in Eq.~(\ref{eq:NOONsimplest_phase}) indicates a combined two-photon phase, $ 2\Theta(\mathbf{r}) \equiv \theta(\mathbf{r}) + \theta(\mathbf{r}') \approx 2\theta(\mathbf{r}) \approx 2\theta(\mathbf{r}')$. For clarity reasons, here we neglect the spatially dependent amplitude, which follows the shape of the normalized irradiance.

\section{Alignment of $HH$ and $VV$ photon pairs} \label{sect:alignment}

\begin{figure}[htb]
    \centering
    \includegraphics[width=0.8\textwidth]{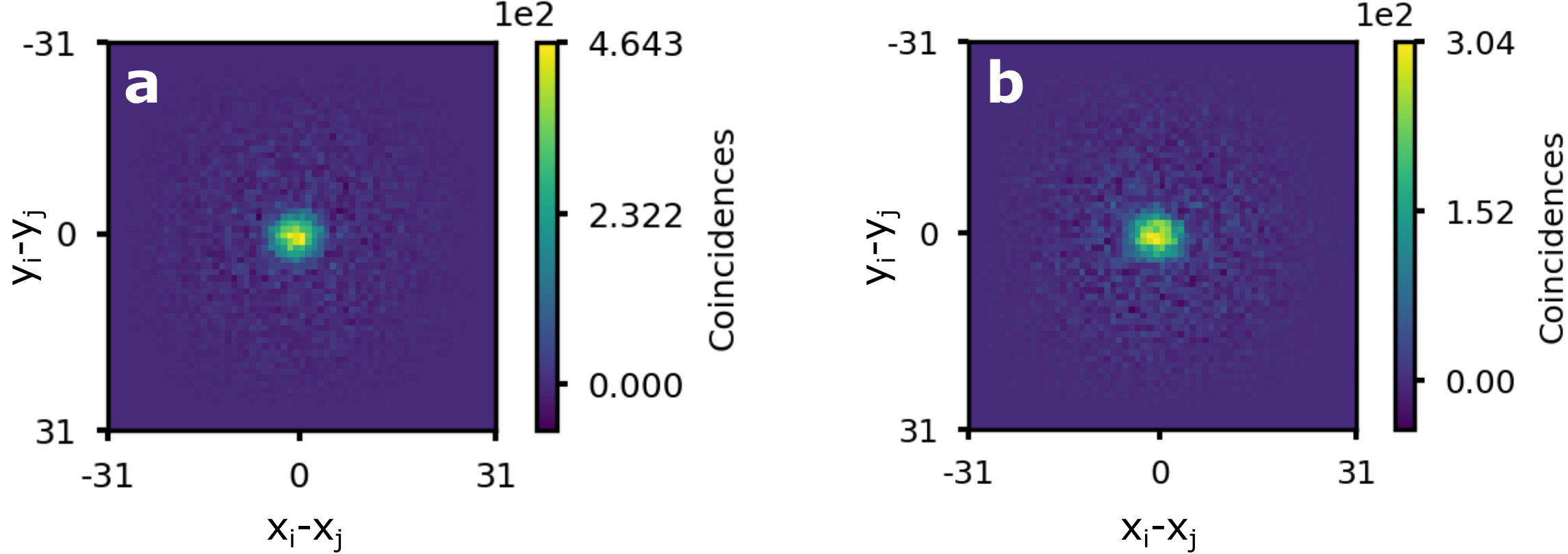}
    \caption{Alignment of position correlations for photon pairs from $HH$ and $VV$ generations. \textbf{(a)} $HH$ generation photon coincidences projected into the difference coordinates $\mathbf{r}_i - \mathbf{r}_j$. Fitted correlation width $\sigma_{-,HH} = 264 \pm 1 \mu m$. \textbf{(b)} $VV$ generation photon coincidences projected into the difference coordinates $\mathbf{r}_i - \mathbf{r}_j$. Fitted correlation width $\sigma_{-,VV} = 275 \pm 2 \mu m$.}
    \label{fig:NF_HH-VV}
\end{figure}

We aligned the $HH$ and $VV$ photon pair generations of our entangled photon source such that they are indistinguishable in position and momentum correlations, which is the crucial requirement for the derivation shown in Section \ref{sect:derivEq1}. The position correlations are measured in a discretized space determined by the SPAD camera resolution, where we consider $\mathbf{r}\longrightarrow \mathbf{r}_{i}$ and $\mathbf{r}'\longrightarrow \mathbf{r}_{j}$. By projecting the coincidences measured in the near-field into the difference coordinates $\mathbf{r}_i - \mathbf{r}_j$, in the presence of SPDC position correlations one obtains a Gaussian peak with waist $\sigma_-$, as can be seen from Eq.~(\ref{eq:SPDC_space}). Through iterative alignment we ensure that the correlation widths for the $HH$ and $VV$ generations are equal, as can be seen in Fig.~\ref{fig:NF_HH-VV}.
We confirm that the correlations widths are matching by fitting a two-dimensional Gaussian function to the difference coordinates projected coincidences, obtaining $\sigma_{-,HH} = 264 \pm 1 \mu m$ for the $HH$ photon pairs, and $\sigma_{-,VV} = 275 \pm 2 \mu m$ for the $VV$ photon pairs (uncertainties are fitting errors). Note that in Fig.~\ref{fig:NF_HH-VV} it can be seen that the amplitudes of the $HH$ and $VV$ generations are different, which we attribute to polarization-dependent losses in optical components of our system. We compensate for this effect however, by increasing the laser pump power for one generation with respect to the other.

\begin{figure}[htb]
    \centering
    \includegraphics[width=\textwidth]{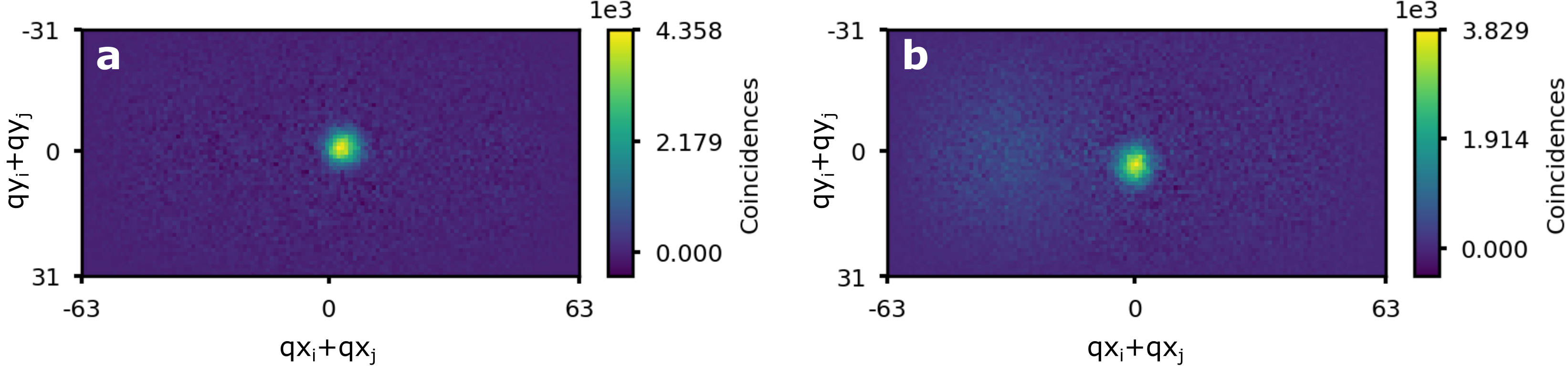}
    \caption{Alignment of momentum correlations for photon pairs from $HH$ and $VV$ generations. \textbf{(a)} $HH$ generation photon coincidences projected into the transverse momentum sum coordinates $\mathbf{q}_i + \mathbf{q}_j$. Fitted correlation width $\sigma_{+,HH}^{-1} = 311 \pm 1 \mu m$. \textbf{(b)} $VV$ generation photon coincidences projected into the transverse momentum sum coordinates $\mathbf{q}_i + \mathbf{q}_j$. Fitted correlation width $\sigma_{+,VV}^{-1} = 326 \pm 2 \mu m$. These images were acquired during the alignment process, ensuring equal momentum correlation widths but prior to ensuring spatial overlap of the $HH$ and $VV$ emissions. This accounts for the relative displacement between the peaks.}
    \label{fig:FF_HH-VV}
\end{figure}

We likewise characterized the momentum correlations of the $HH$ and $VV$ generations from Eq.~(\ref{eq:gamma_momentum}), by considering $\mathbf{q}\longrightarrow\mathbf{q}_{i}$ and $\mathbf{q}'\longrightarrow\mathbf{q}_{j}$ and mapping the two-photons momenta ($\mathbf{q}_i + \mathbf{q}_j$) from the detected lateral position coordinates measured in the far-field.  Accordingly, one obtains a Gaussian peak with waist $\sigma_{+}^{-1}$, the momentum correlation width. We again aligned the $HH$ and $VV$ generations such that their momentum correlations were equal, which is shown in Fig.~\ref{fig:FF_HH-VV}. Through an iterative alignment procedure, we moreover ensured that the $HH$ and $VV$ generations had equal correlation widths for \textit{both} position and momentum correlations. We confirmed that the momentum correlation widths are matching again by fitting a two-dimensional Gaussian function, where we obtained $\sigma_{+,HH}^{-1} = 311 \pm 1 \mu m$ for the $HH$ photon pairs, and $\sigma_{+,VV}^{-1} = 326 \pm 2 \mu m$ for the $VV$ photon pairs (uncertainties are fitting errors). After satisfying the requirement of equal correlation widths in position and momentum, we finally ensured excellent spatial overlap of the intensity profiles of the two generations, which completed the alignment.

\section{Details of coincidence imaging}

\subsection{Crosstalk probabilities}
\begin{figure}[htb]
    \centering
    \includegraphics[width=0.7\textwidth]{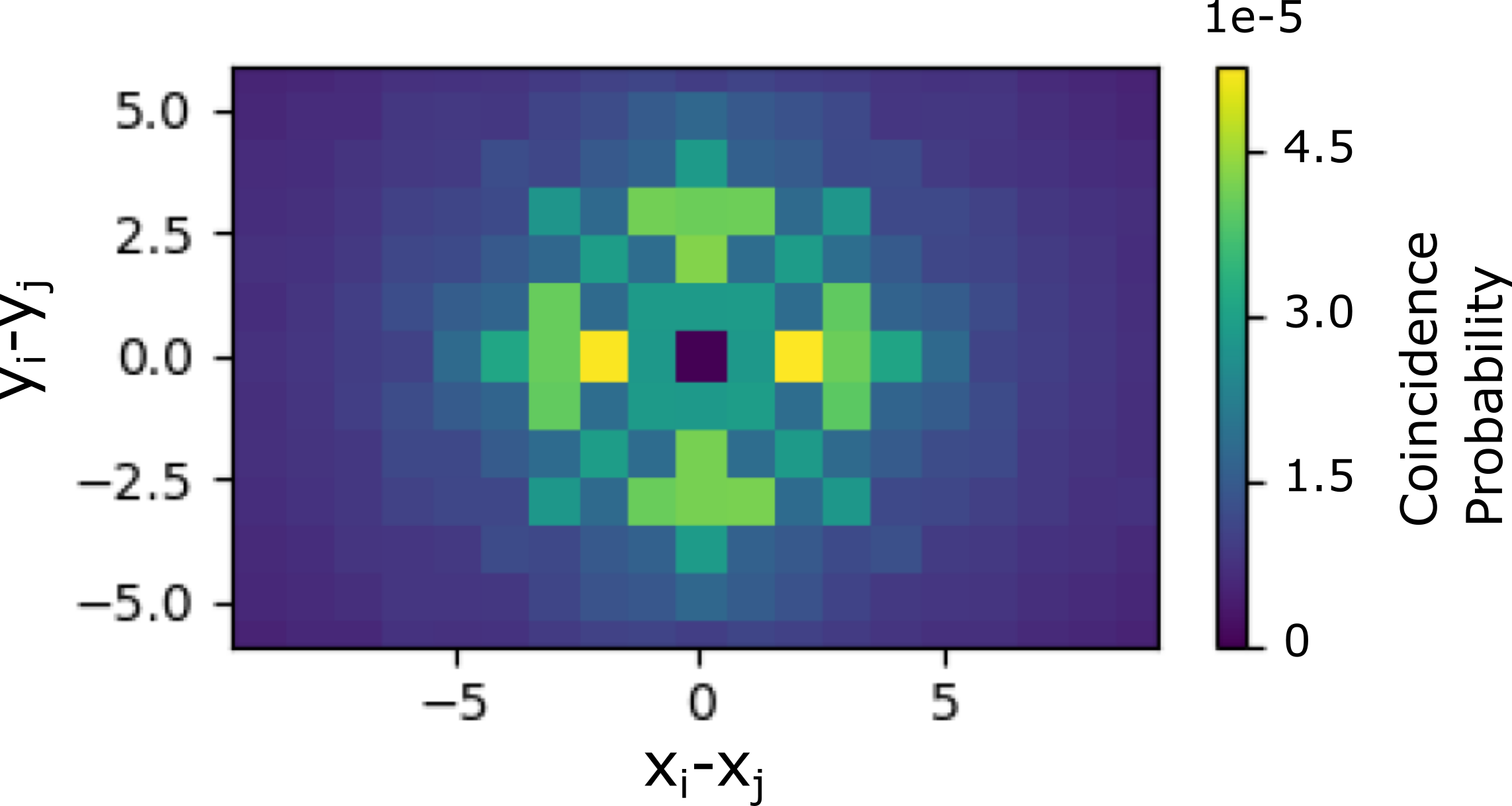}
    \caption{Probability of crosstalk coincidences as a function of pixel displacement.}
    \label{fig:xtalk}
\end{figure}

The map of crosstalk coincidence probabilities $P_\mathrm{ct}$ as a function of pixel displacement (acquisition described in Methods of main article) is shown in Fig.~\ref{fig:xtalk}. We show only the central region of the probability map with pixel displacement close to zero, as the crosstalk probability is negligible for pixel pairs that are further separated. Note also that $P_\mathrm{ct}$ depends only on $\Delta x = |x_i - x_j|$ and $\Delta y = |y_i - y_j|$, and that the probability map is therefore symmetric about zero, i.e. $P_\mathrm{ct}(x_i-x_j,y_i-y_j) = P_\mathrm{ct}(x_j-x_i,y_i-y_j) = P_\mathrm{ct}(x_i-x_j,y_j-y_i) = P_\mathrm{ct}(x_j-x_i,y_j-y_i)$.

\subsection{Uncorrelated noise removal}

We attribute spatially uncorrelated noise coincidence events to for example stray light, spurious reflections in the setup, and imperfections in the nonlinear crystal generating SPDC photon pairs that are not spatially correlated. Here we describe the fitting procedure used to filter out such spatially uncorrelated noise coincidences. We begin by noting that for a fixed given pixel $i$ with coordinates $[x_i, y_i]$, the set of possible coincidences with all other pixels forms a two-dimensional image. This can be denoted as $cc(x,y|x_i,y_i)$, i.e. the coincidence counts conditional on a detection having been registered at pixel $i$. For every $i^{th}$ pixel we fit $cc(x,y|x_i,y_i)$ with a two-dimensional Gaussian model which has the form
\begin{equation}
\label{eq:gaussianFit}
    G_i(x, y) = A_i \, \exp \left[ \frac{-(((x-x_{0,i})^2 +(y-y_{0,i})^2)}{2 \sigma_{\mathrm{fit},i}^2} \right],
\end{equation}
where $A_i$ is an amplitude fitting parameter, $[x_{0,i}, y_{0,i}]$ is the peak location, and $\sigma_{\mathrm{fit},i}$ is the fitted waist. Note that for coincidence counts acquired in the $\bra{DD}$ and $\bra{AA}$ projections ($cc_{DD}$ and $cc_{AA}$), we have $[x_{0,i}, y_{0,i}] = [x_i, y_i]$. This is not the case for coincidence counts acquired in the $\bra{DA}$ polarization projection ($cc_{DA}$) where the two photons are detected on different halves of the camera. This is the reason why the $x_{0,i}$ and $y_{0,i}$ fitting parameters are necessary in Eq.~(\ref{eq:gaussianFit}). The width of the Gaussian fitting used for filtering is then obtained by taking the mean of all fitted $\sigma_{\mathrm{fit},i}$ values. We further calculate the offsets between the Gaussian peak $[x_{0,i}, y_{0,i}]$ and the pixel $[x_i, y_i]$ for the $cc_{DA}$ coincidences $d_{x,i} = x_{0,i} - x_i$ and $d_{y,i} = y_{0,i} - y_i$. The offset values $d_x$ and $d_y$ used in the Gaussian fitting for filtering are again obtained by taking the mean of all $d_{x,i}$ and $d_{y,i}$ values (averaged over all pixels $i$). Here we assume that the photon pair correlations are uniform across the camera sensor, and that differences in $\sigma_{\mathrm{fit},i}$, and $d_{x,i}$ and $d_{y,i}$, between different $i^{th}$ pixels are simply due to local variations in photon counting statistics. 

\begin{figure}[htb]
    \centering
    \includegraphics[width=\textwidth]{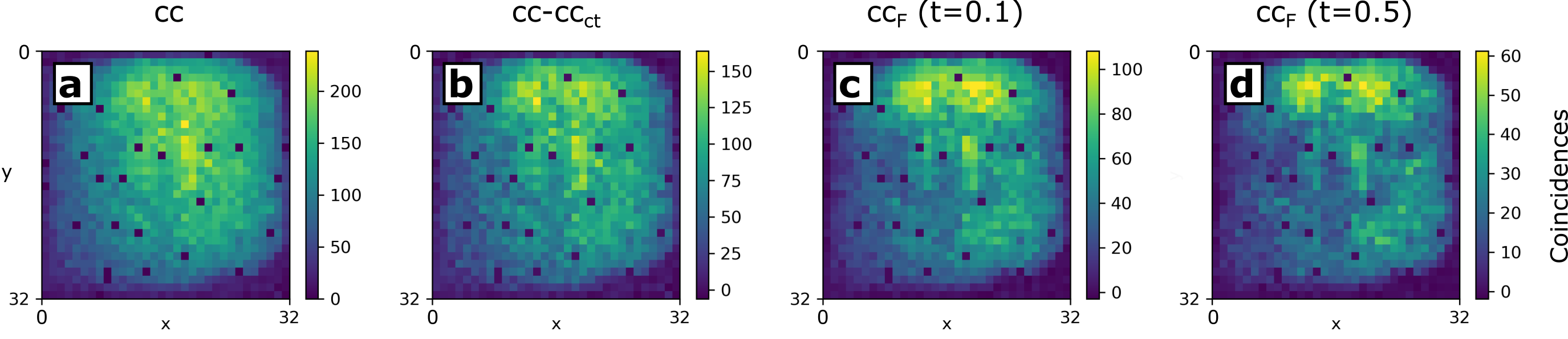}
    \caption{\textbf{(a)} Coincidence image with no noise removed. \textbf{(b)}  Coincidence image with crosstalk removed. \textbf{(c)} Coincidence image with crosstalk removed and filtering (threshold $t=0.1$) to remove spatially uncorrelated noise. \textbf{(d)} Coincidence image with crosstalk removed and filtering (threshold $t=0.5$) to remove spatially uncorrelated noise. }
    \label{fig:cc_filtering}
\end{figure}

The influence on a coincidence image of crosstalk reduction and coincidence filtering for uncorrelated noise removal are shown in Fig.~\ref{fig:cc_filtering}. In particular, it can be seen that for a higher filtering threshold $t$, the spatial resolution improves, however this comes at the expense of counts detected. We therefore chose a filtering threshold of $t=0.5$ to achieve a good spatial resolution without losing an excessive number of detections.

\subsection{Mapping coincidence counts onto images}

Eq.~(2) in the main text of this article describes the mapping of coincidence counts (a four-dimensional quantity - two spatial dimensions per pixel) onto a two-dimensional image. This transformation relies on the photon pairs being spatially correlated at the point of detection, i.e. being approximately in the same position. If this condition is satisfied, $cc(x,y|x_i,y_i)$ the set of coincidences with a given pixel $[x_i, y_i]$, describes a point-like image with the width of the point given by the photon pair correlation width, which can be integrated over to obtain the two-photon counts at $[x_i, y_i]$.

\begin{figure}
    \centering
    \includegraphics[width=\textwidth]{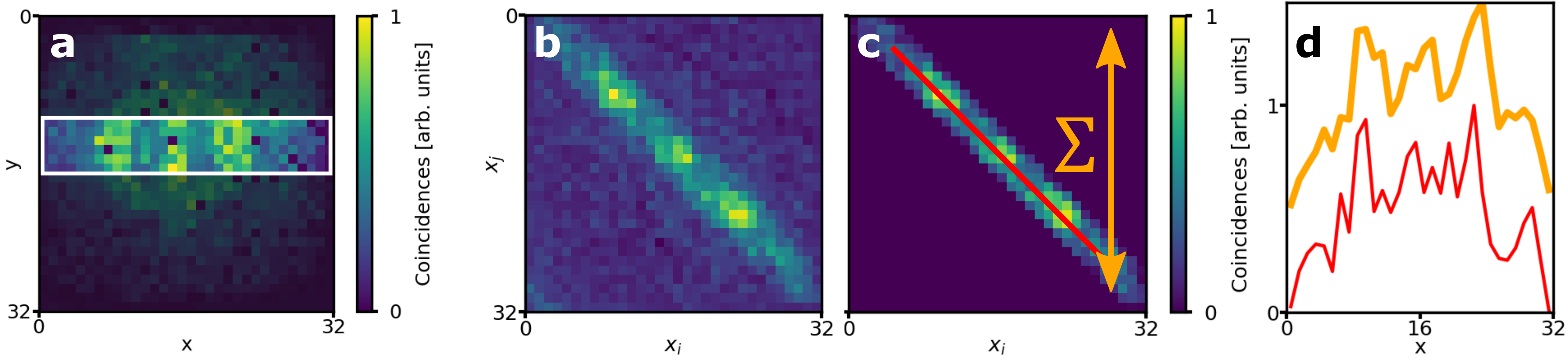}
    \caption{Mapping coincidence matrix for two photons in $x$ onto two-photon counts vector in $x$. \textbf{(a)} Quasi one-dimensional section of coincidence image: the section highlighted by the white rectangle can be approximately described using only the $x$ coordinate of photon detections. \textbf{(b)} Coincidence matrix for coincidences between photon $i$ detected at $x_i$ and photon $j$ detected at $x_j$. \textbf{(c)} Coincidence matrix from \textbf{(b)} filtered to remove spatially uncorrelated noise. \textbf{(d)} Two-photon counts as a function of $x$. Red curve: normalized diagonal values from coincidence matrix (red line in \textbf{(c)}). Orange curve: normalized two-photon counts obtained by integrating over $x_j$ in \textbf{(c)}. Orange curve is displaced vertically for clarity.}
    \label{fig:coincMapping1D}
\end{figure}

This process can be appreciated more intuitively by considering coincidences in one spatial dimension. Let us consider a phase pattern approximately invariant in the $y$-axis, for example the highlighted region within the white rectangle in Fig.~\ref{fig:coincMapping1D}a. For the equivalent one-dimensional profile in the $x$-axis, one can register and visualize the associated coincidences between one photon at location $x_i$ and another one at any other location $x_j$ as shown in Fig.~\ref{fig:coincMapping1D}b, where here we show coincidences measured in the $\bra{DA}$ polarization projection. This pattern is a two-dimensional coincidence map, where spatially correlated photon pairs, like the ones from the near-field of our N00N state, are located around the diagonal elements. Spurious coincidences, which are spatially uncorrelated, represent background noise and can be filtered as shown in Fig.~\ref{fig:coincMapping1D}c, according to Eq.~(9) in the main text of this article. Here, we only keep photon-coincidences which comes from detections that were close in space. The diagonal elements (red line in Fig.~\ref{fig:coincMapping1D}c) represent the two-photon detections for exactly each spatial coordinate $x$. However, plotting this (red curve in Fig.~\ref{fig:coincMapping1D}d) reveals that the diagonal (normlized) two-photon counts are quite noisy, as it is relatively unlikely for both photons to be in precisely the same location. Fortunately, when the two photons are strongly spatially correlated, the photon detections which lie close to the diagonal, i.e. which satisfy $x_{i} - \epsilon < x_{j} < x_{i} + \epsilon$ (where $\epsilon$ is the width in pixels of the filtering as applied by Eq.~(9) from the main text), carry approximately the same information as the diagonal components. Our protocol therefore extracts equivalent information by summing the contributions along one axis of the elements of the filtered coincidence map ($x_{i}\approx x_{j}$), as indicated by the orange arrow in Fig.~\ref{fig:coincMapping1D}c. Consequently, a one-dimensional profile of two-photon counts is retrieved (orange curve in Fig.~\ref{fig:coincMapping1D}d), with the same shape as the diagonal but with less noise.

For two-dimensional phase profiles, the coincidence map lives in a four-dimensional space, where coincidences are combined in two axes rather than one. The two-dimensional coincidence image is then computed by using Eq.~(2) in the main text.

\section{Classical and N00N state interference}

Classical interference images were obtained by using a single SPDC generation as a source of low-intensity light. H polarized photons were generated by our SPDC source, and rotated with a HWP into the diagonal basis. As with polarization entangled state, the total phase $\Theta(\mathbf{r})$ represents both birefringent or non-birefringent phase as well as the LIM controlled bias phase. For a given $\Theta(\mathbf{r})$, there are two possible outcomes after projecting into either the diagonal or anti-diagonal polarization before imaging. Concretely, on the left half of the SPAD array camera the detection probability
\begin{equation}
\label{eq:classInterferenceD}
    P_{\mathrm{Classical, D}} = \bra{D}\left(\frac{\ket{H(\mathbf{r})}+e^{i\Theta(\mathbf{r})}\ket{V(\mathbf{r})}}{\sqrt{2}}\right) = \frac{\mathcal{V}_{\mathrm{Classical}}\cos(\Theta(\mathbf{r})) + 1}{2},
\end{equation}
whereas the right side of the SPAD camera measures an intensity with a detection probability of
\begin{equation}
    \label{eq:classInterferenceA}
    P_{\mathrm{Classical, A}} = \bra{A}\left(\frac{\ket{H(\mathbf{r})}+e^{i\Theta(\mathbf{r})}\ket{V(\mathbf{r})}}{\sqrt{2}}\right) = \frac{-\mathcal{V}_{\mathrm{Classical}}\cos(\Theta(\mathbf{r})) + 1}{2}
\end{equation}
where $\mathcal{V}_{\mathrm{Classical}}$ is the classical interference visibility. The actual detected intensities $I$ are obtained simply by multiplying these probabilities by the number of photons used. See ({\it 17\/}) for further details on the derivation of Eq.~(\ref{eq:classInterferenceD}) and \ref{eq:classInterferenceA}. In our case we obtained very high classical interference visibilities $\mathcal{V}_{\mathrm{Classical}}$ close to unity.

On the other hand, for a polarization N00N state (as described by Eq.~(\ref{eq:NOONsimplest_phase})) a phase between the $HH$ and $VV$ components is transformed into a measurable change in coincidence counts by projecting into one of the three diagonal two-photon bases. This yields the following coincidence detection probabilities:
\begin{align}
    P_{\mathrm{N00N, DD}} = \braket{DD|\psi(\mathbf{r})} &= \frac{\mathcal{V}\cos(2 \Theta(\mathbf{r})) + 1}{4}
    \label{eq:NOONinterference_DD} \\
    P_{\mathrm{N00N, AA}} = \braket{AA|\psi(\mathbf{r})} &= \frac{\mathcal{V}\cos(2 \Theta(\mathbf{r})) + 1}{4}
    \label{eq:NOONinterference_AA} \\
    P_{\mathrm{N00N, DA}} = \braket{DA|\psi(\mathbf{r})} &= \frac{-\mathcal{V}\cos(2 \Theta(\mathbf{r})) + 1}{2}
    \label{eq:NOONinterference_DA}
\end{align}
where $ D \equiv (H+V)/\sqrt{2}$ and $ A \equiv (H-V)/\sqrt{2}$, and $\mathcal{V}$ is the N00N state interference visibility. Note that as in our work the two photons are indistinguishable, the $\braket{DA|\psi}$ projective measurement is indistinguishable from the $\braket{AD|\psi}$ projection, which accounts for Eq.~(\ref{eq:NOONinterference_DA}) having twice the amplitude as Eq.~(\ref{eq:NOONinterference_DD}) and Eq.~(\ref{eq:NOONinterference_AA}). The actual detected coincidence counts $cc$ (which are used to form the coincidence images $ci$) are obtained simply by multiplying these probabilities by the number of entangled photon pairs used. The N00N state interference visibility $\mathcal{V}$ is characterized in the main body of this article, where we obtained a fitted visibility of $0.94\pm0.06$.

\section{Sensitivity enhancement with N00N state phase imaging}

\subsection{Noise of coincidence counting} \label{sect:cc_noise}

In the ideal case of shot noise limited measurements the standard deviation of the coincidence images and intensity images measured is simply given by the shot noise, i.e. $\mathrm{sd}(ci)_{\mathrm{ideal}} = \sqrt{ci}$ and $\mathrm{sd}(I)_{\mathrm{ideal}} = \sqrt{I}$. However, the coincidence counting method used in this work can decrease the SNR of measured coincidence count. We therefore characterize the noise which is added by our coincidence counting method, in order to ensure that super-sensitive phase measurements can be achieved with the PSDH method. We denote the measured standard deviation of $ci$ as $\mathrm{sd}(ci)_{\mathrm{meas}} = \kappa \sqrt{ci}$, where $\kappa \geq 1$ is the increase in noise caused by our coincidence counting method. For the classical case, we always assume shot noise limited measurements in order to theoretically compare our quantum-enhanced method against the best possible classical measurement.

\begin{figure}[htb]
    \centering
    \includegraphics[width=0.6\textwidth]{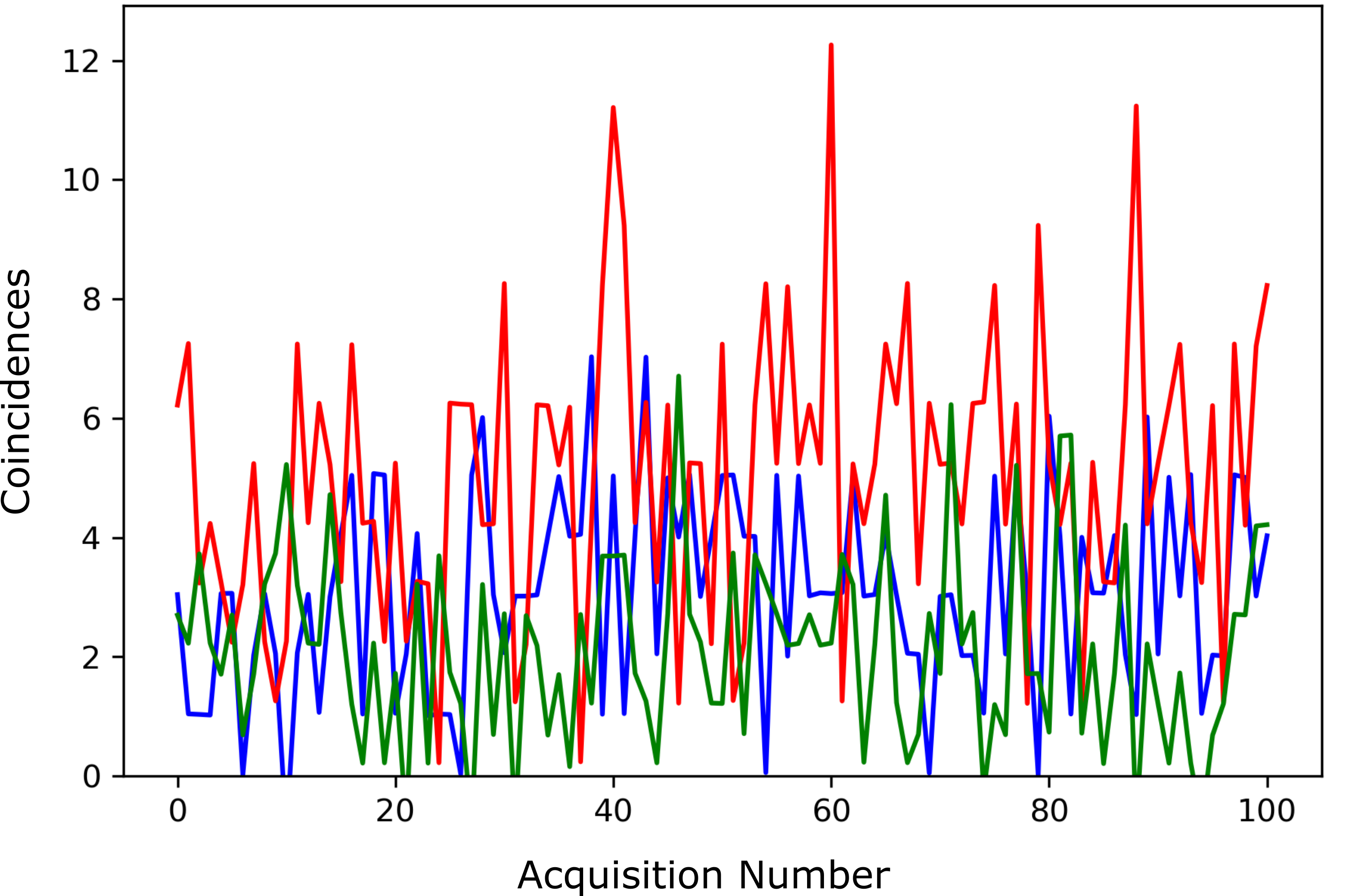}
    \caption{Variation in detected coincidences. Red: DA coincidences, Blue: DD coincidences, Green: AA coincidences.}
    \label{fig:cc_0.62mW}
\end{figure}

\begin{figure}[htb]
    \centering
    \includegraphics[width=\textwidth]{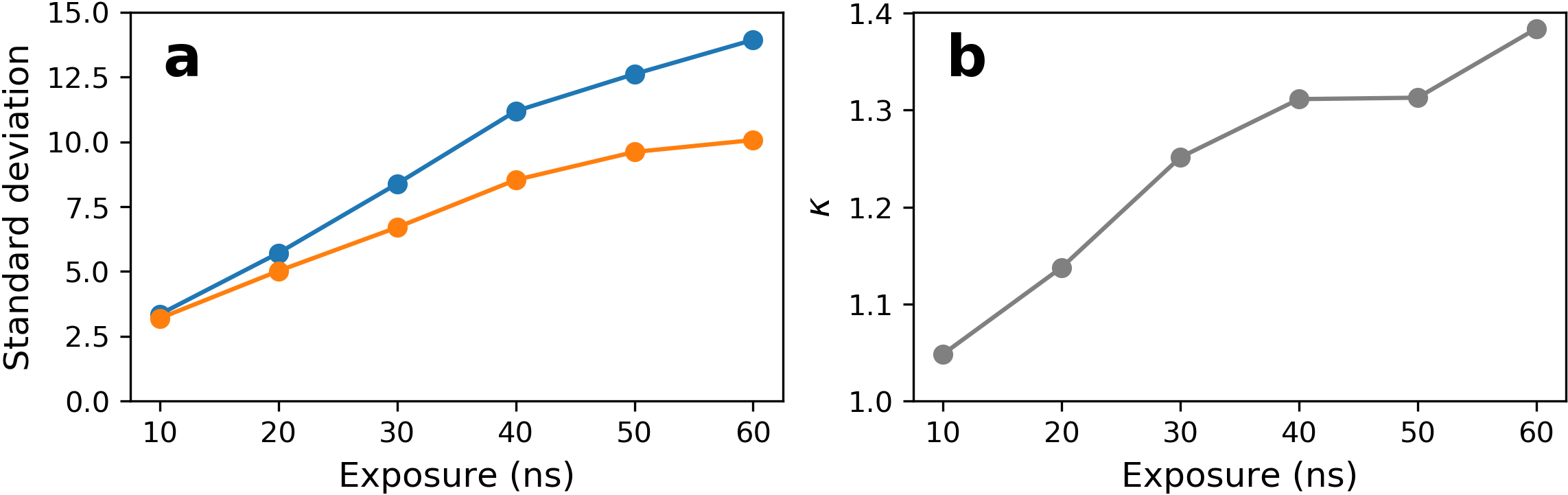}
    \caption{Noise added by coincidence counting method. \textbf{(a)} Blue: experimentally measured standard deviation of coincidence counts $\mathrm{sd}(ci_i)_{\mathrm{meas}}$. Orange: Ideal (shot noise limited) standard deviation of coincidence counts $\mathrm{sd}(ci_i)_{\mathrm{ideal}}$. \textbf{(b)} Increase in noise of coincidence counts $\kappa$ (compared to ideal case).}
    \label{fig:exposureVsStats}
\end{figure}

A full theoretical treatment of the increase in noise $\kappa$ can be found in ({\it 24\/}). In this work we experimentally measured $\kappa$ of our coincidence measurements by uniformly illuminating the SPAD camera with our entangled photon source, and acquiring a series of many coincidence acquisitions, each with acquisition time \SI{5}{\second}. One example of such an acquisition series is shown in Fig.~\ref{fig:cc_0.62mW}, where in this case the laser pump power for the SPDC photon source was \SI{0.6}{\milli\watt}, and the exposure time per frame was \SI{10}{\nano\second}. Coincidence images were acquired and the counts integrated over the whole camera sensor. The experimentally measured standard deviation $\mathrm{sd}(ci)_{\mathrm{meas}}$ was calculated simply taking the standard deviation of the counts over the acquisition series. The ideal (shot noise limited) standard deviation $\mathrm{sd}(ci)_{\mathrm{ideal}}$ on the other hand is simply the square root of the mean number of counts over the acquisition series. The ideal and measured standard deviations are plotted, for a pump laser power of \SI{0.6}{\milli\watt} and different exposure times, in Fig.~\ref{fig:exposureVsStats}a. We then have $\kappa = \mathrm{sd}(ci)_{\mathrm{meas}} / \mathrm{sd}(ci)_{\mathrm{ideal}}$, which is plotted in Fig.~\ref{fig:exposureVsStats}b. In particular, it can be seen that for an exposure time of \SI{10}{\nano\second}, $\kappa = 1.05$ which is close to unity meaning that under these experimental conditions our coincidence counting method operates close to the ideal shot noise limit. We therefore used these experimental conditions when imaging samples in this work.

\subsection{Noise of PSDH-retrieved phase estimates}

Recall that using classical and N00N state interference, an estimate of the sample phase can be retrieved using phase-shifting digital holography (PSDH). As is detailed in the main body of this article, for classical interference resulting in intensity images this is calculated according to the formula
\begin{equation}
    \label{eq:PSDH_intensity_SM}
    \hat{\phi}_{\mathrm{Classical}}(\mathbf{r}) = \tan^{-1}\left[\frac{I(\phi_{\mathrm{Sample}}(\mathbf{r}), \pi/2) - I(\phi_{\mathrm{Sample}}(\mathbf{r}), 3\pi/2)}{I(\phi_{\mathrm{Sample}}(\mathbf{r}), \pi) - I(\phi_{\mathrm{Sample}}(\mathbf{r}), 0)}\right].
\end{equation}
On the other hand, for N00N state interference resulting in coincidence images, this is calculated according to the formula
\begin{equation}
\label{eq:PSDH_NOON_SM}
    \hat{\phi}_{\mathrm{N00N}}(\mathbf{r}) = \frac{1}{2} \tan^{-1}\left[\frac{ci(\phi_{\mathrm{Sample}}(\mathbf{r}),\pi/4) - ci(\phi_{\mathrm{Sample}}(\mathbf{r}), 3\pi/4)}{ci(\phi_{\mathrm{Sample}}(\mathbf{r}), \pi/2) - ci(\phi_{\mathrm{Sample}}(\mathbf{r}), 0)}\right].
\end{equation}
For conciseness in this section we henceforth denote the intensity images $I(\phi_{Sample}(\mathbf{r}),\alpha)$, where $\alpha = \{0, \pi/2, \pi, 3\pi/2\}$, as $I_j$, with $j = {0,1,2,3}$. Similarly, we denote the four coincidence image variables $ci(\phi_{Sample}(\mathbf{r}),\alpha)$, where $\alpha = \{0, \pi/4, \pi/2, 3\pi/4\}$, as $ci_j$, with $j = {0,1,2,3}$. Note that $ \hat{\phi}_{\mathrm{Classical}} $ and $ \hat{\phi}_{\mathrm{N00N}} $ are acquired for each of the polarization projection measurements $D$ and $A$ (for classical), and $DD$, $AA$ and $DA$ (for N00N). The phase estimates acquired using the different polarization projections are then combined as follows
\begin{align}
    \hat{\phi}_{\mathrm{Classical}} & = \frac{\hat{\phi}_{\mathrm{Classical, D}} + \hat{\phi}_{\mathrm{Classical, A}}}{2} \label{eq:combinephi_classical_SM}\\
    \hat{\phi}_{\mathrm{N00N}} & = \frac{\hat{\phi}_{\mathrm{N00N, DD}} + \hat{\phi}_{\mathrm{N00N, AA}} + 2\hat{\phi}_{\mathrm{N00N, DA}}}{4}, \label{eq:combinephi_NOON_SM}
\end{align}
where the factor of two of the $DA$ term in Eq.~(\ref{eq:combinephi_NOON_SM}) is due to the indistinguishability of the $\bra{DA}$ and $\bra{AD}$ polarization measurements (as in Eq.~(\ref{eq:NOONinterference_DA})).

Uncertainty, i.e. noise, in the phase estimates results from the noise in the experimentally measured $I_j$ and $ci_j$ quantities, as detailed in Section \ref{sect:cc_noise}. Substituting Eq.~(\ref{eq:classInterferenceD})-(\ref{eq:NOONinterference_DA}), we obtain the following expressions for the standard deviations of the $I_j$ measurables:
\begin{equation}\label{eq:sd_I_SM}
    sd(I_j) = 
    \begin{cases}
    & \left[\frac{\mathcal{V}_{\mathrm{Classical}}\cos[\phi_{\mathrm{Sample}}(\mathbf{r}) + j\pi/2] + 1}{2} I_{\mathrm{tot}}/4 \right]^{1/2} \quad  \text{for $I_D$,}\\
    & \left[\frac{-\mathcal{V}_{\mathrm{Classical}}\cos[\phi_{\mathrm{Sample}}(\mathbf{r}) + j\pi/2] + 1}{2} I_{\mathrm{tot}}/4 \right]^{1/2} \quad  \text{for $I_A$,}
    \end{cases}
\end{equation}
where $I_{tot}$ is the total number of photons used over all four acquisition steps, i.e. $I_{tot}/4 = I_{j,D} + I_{j,A}$ is the number of photons used for each $j^{th}$ PSDH acquisition step. Likewise for the $ci_j$ measurables we have:
\begin{equation}\label{eq:sd_ci_SM}
    sd(ci_j) = 
    \begin{cases}
    & \kappa \big[\frac{\mathcal{V}\cos\left[2\phi_{\mathrm{Sample}}(\mathbf{r}) + j\pi/4\right] + 1}{4} ci_{\mathrm{tot}}/4 \big]^{1/2} \quad  \text{for $cc_{DD}$ and $cc_{AA}$,}\\
    & \kappa \big[\frac{-\mathcal{V}\cos\left[2\phi_{\mathrm{Sample}}(\mathbf{r}) + j\pi/4\right] + 1}{2} ci_{\mathrm{tot}}/4 \big]^{1/2} \quad  \text{for $cc_{DA}$,}
    \end{cases}
\end{equation}
where $ci_{tot}$ is the total number of photon coincidences (i.e. entangled photon pairs) used over all four acquisition steps, i.e. $ci_{tot}/4 = ci_{j,DD} + ci_{j,AA} + ci_{j,DA}$ is the total number of coincidences for each $j^{th}$ PSDH acquisition step. Recall that in both Eq.~(\ref{eq:sd_I_SM}) and Eq.~(\ref{eq:sd_ci_SM}) we have $j = {0,1,2,3}$. In order to relate the standard deviations of $I_j$ and $ci_j$ to the standard deviations of the retrieved phase estimates, we use the error propagation formula ({\it 50\/}):
\begin{equation}
    \label{eq:errorProp_SM}
    \mathrm{sd}(f) = \sqrt{\sum_j \left[ \left(\frac{\partial f}{\partial x_j}\right)^2 \mathrm{sd}(x_j)^2 \right]},
\end{equation}
where in general $f$ is a function that depends on multiple variables $x_j$. In the case of N00N(classical) interference measurements, the function $f$ is $\hat{\phi}_{\mathrm{N00N}}$($\hat{\phi}_{\mathrm{Classical}}$) which depends on the experimentally measured $ci$($I$) variables. The partial derivates $\partial f/\partial x_j$ are then:
\begin{align}
    \frac{\partial \hat{\phi}_{\mathrm{Classical}}}{\partial I_0} &= \frac{I_1 -I_3}{(I_0 - I_2)^2 + (I_1 - I_3)^2} \label{eq:partialI_first} \\
    \frac{\partial \hat{\phi}_{\mathrm{Classical}}}{\partial I_1} &= \frac{I_2 -I_0}{(I_0 - I_2)^2 + (I_1 - I_3)^2} \\
    \frac{\partial \hat{\phi}_{\mathrm{Classical}}}{\partial I_2} &= \frac{I_3 -I_1}{(I_0 - I_2)^2 + (I_1 - I_3)^2} \\
    \frac{\partial \hat{\phi}_{\mathrm{Classical}}}{\partial I_3} &= \frac{I_0 -I_2}{(I_0 - I_2)^2 + (I_1 - I_3)^2} \label{eq:partialI_last}
\end{align}
and
\begin{align}
    \frac{\partial \hat{\phi}_{\mathrm{N00N}}}{\partial ci_0} &= \frac{ci_1 -ci_3}{2((ci_0 - ci_2)^2 + (ci_1 - ci_3)^2)} \label{eq:partialci_first} \\
    \frac{\partial \hat{\phi}_{\mathrm{N00N}}}{\partial ci_1} &= \frac{ci_2 -ci_0}{2((ci_0 - ci_2)^2 + (ci_1 - ci_3)^2)} \\
    \frac{\partial \hat{\phi}_{\mathrm{N00N}}}{\partial ci_2} &= \frac{ci_3 -ci_1}{2((ci_0 - ci_2)^2 + (ci_1 - ci_3)^2)} \\
    \frac{\partial \hat{\phi}_{\mathrm{N00N}}}{\partial ci_3} &= \frac{ci_0 -ci_2}{2((ci_0 - ci_2)^2 + (ci_1 - ci_3)^2)}. \label{eq:partialci_last} 
\end{align}
To obtain the standard deviation of the final retrieved phase estimate, combined from the different polarization measurements, one again applies Eq.~(\ref{eq:errorProp_SM}) to Eq.~(\ref{eq:combinephi_classical_SM}) and Eq.~(\ref{eq:combinephi_NOON_SM}).

\begin{figure}[htb]
    \centering
    \includegraphics[width=\textwidth]{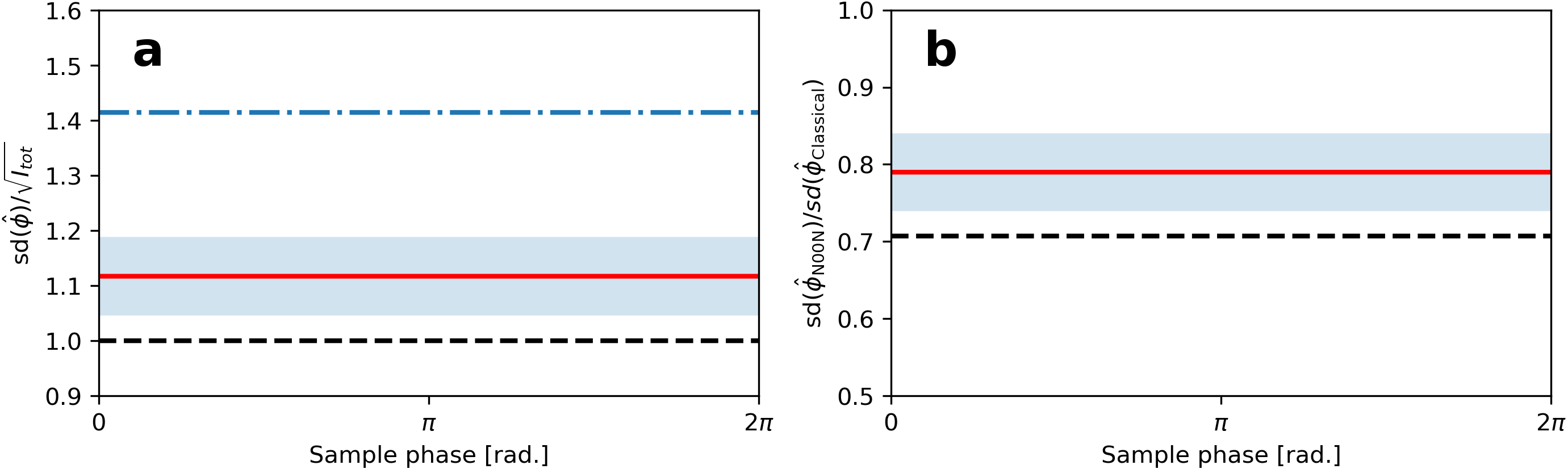}
    \caption{Standard deviations of entanglement-enhanced phase estimates vs classical phase estimates. \textbf{(a)} Normalized standard deviations of phase estimates $sd(\hat{\phi})/\sqrt{I_{tot}}$, as a function of the sample phase. Blue dotted dashes: classical measurement. Red solid line: entanglement-enhanced measurement with $\kappa = 1.05$, $\mathcal{V}= 0.94$. Shaded area: uncertainty around red line, caused by visibility ($\mathcal{V}$) uncertainty of $\pm0.06$. Black dashed line: ideal entanglement-enhanced measurement. \textbf{(b)} Reduction in noise with entanglement-enhanced vs ideal classical measurement. Red solid line: noise reduction due to entanglement-enhanced measurement with $\kappa = 1.05$, $\mathcal{V}= 0.94$. Shaded area: uncertainty around red line, caused by visibility ($\mathcal{V}$) uncertainty of $\pm0.06$. Black dashed line: noise reduction due to ideal entanglement-enhanced measurement.}
    \label{fig:SNR_PSDH}
\end{figure}

For a fair sensitivity comparison between the classical and entanglement-enhanced method, we compare the standard deviations of $\hat{\phi}_{\mathrm{Classical}}$ and $\hat{\phi}_{\mathrm{N00N}}$ with an equal number of total photons used, i.e. $ci_{tot} = I_{tot}/2$ (as the coincidence image $ci$ is composed of coincidence, i.e. two-photon, count terms). Therefore substituting Eq.~(\ref{eq:sd_I_SM}) and Eq.~(\ref{eq:sd_ci_SM}) into the error propagation formula Eq.~(\ref{eq:errorProp_SM}), using the partial derivatives Eq.~(\ref{eq:partialI_first})-(\ref{eq:partialI_last}) and Eq.~(\ref{eq:partialci_first})-(\ref{eq:partialci_last}) respectively, we can calculate the normalized standard deviation of the phase estimate $sd(\hat{\phi})/\sqrt{I_{tot}}$ as a function of the sample phase. We calculated the normalized standard deviations $\hat{\phi}_{\mathrm{Classical}}/\sqrt{I_{tot}}$ and $\hat{\phi}_{\mathrm{N00N}}/\sqrt{I_{tot}}$ for the ideal shot-noise limited cases, as well as for the experimentally measured case with $\kappa = 1.05$ and $\mathcal{V} = 0.94 \pm 0.06$. This is shown in Fig.~\ref{fig:SNR_PSDH}a, where we note that the noise (standard deviation) in the PSDH-retrieved phase estimates is constant across all sample phase values. We further calculated the reduction in noise of the phase estimates using our entanglement-enhanced PSDH-retrieved phase estimates as compared to the ideal classical case, which is shown in Fig.~\ref{fig:SNR_PSDH}b. Again, note that the sensitivity improvement (i.e. reduction in noise) is constant across the sample phase values. For the ideal, shot noise limited entanglement-enhanced phase measurement, the reduction in noise corresponds exactly to the expected factor of $1/\sqrt{2}\approx 0.707$ for N00N state phase measurements, whereas for our experimental coincidence measuring conditions the noise is reduced by a factor of $0.789 \pm 0.050$.

\section{Subtraction of background phase profile}

The non-zero background phase profile measured in our system, i.e. the phase that is measured in the absence of any sample, is caused by several factors. Imperfect collimation of the beam through the LIM results in a distortion of the wavefront difference between the $H$ and $V$ polarized photons ({\it 54\/}). We have also observed that various components in the optical path going from the entangled photon source to the LIM, most of all the dichroic mirror, induce small birefringent effects which again cause spatially dependent phase differences between the $H$ and $V$ polarized photons. Lastly, imperfect alignment of the Sagnac interferometer and inhomogeneities in the ppKTP crystal also result in non-zero phase differences between $H$ and $V$.

\begin{figure}[hbt]
    \centering
    \includegraphics[width=0.6\textwidth]{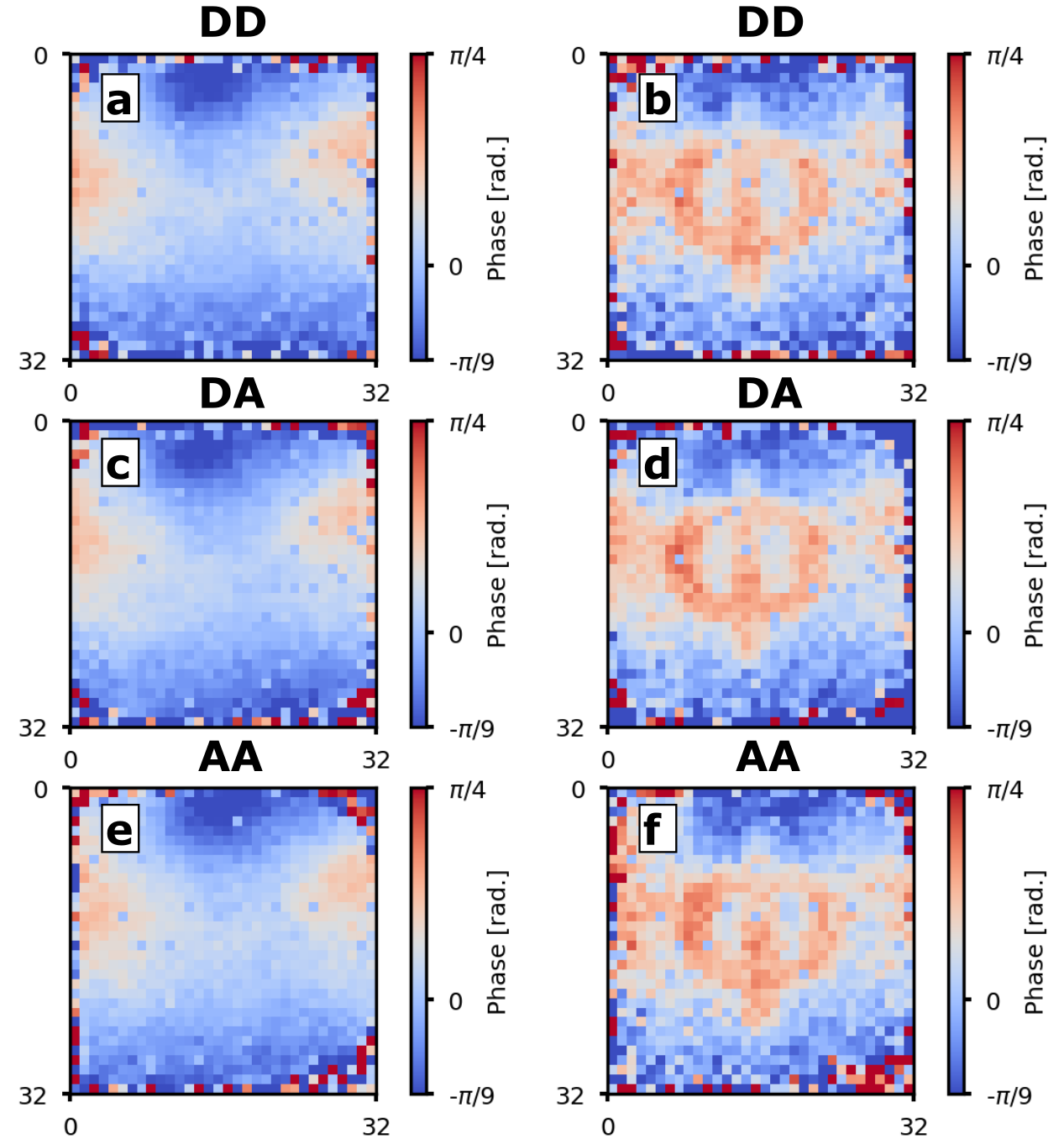}
    \caption{Retrieved phase images for birefringent test sample measurement. \textbf{(a)} and \textbf{(b)} Background and sample retrieved phase for $\bra{DD}$ polarization projection. \textbf{(c)} and \textbf{(d)} Background and sample retrieved phase for $\bra{DA}$ polarization projection. \textbf{(e)} and \textbf{(f)} Background and sample retrieved phase for $\bra{AA}$ polarization projection.}
    \label{fig:backgroundPhaseSubtr_BP}
\end{figure}

\begin{figure}[hbt]
    \centering
    \includegraphics[width=0.6\textwidth]{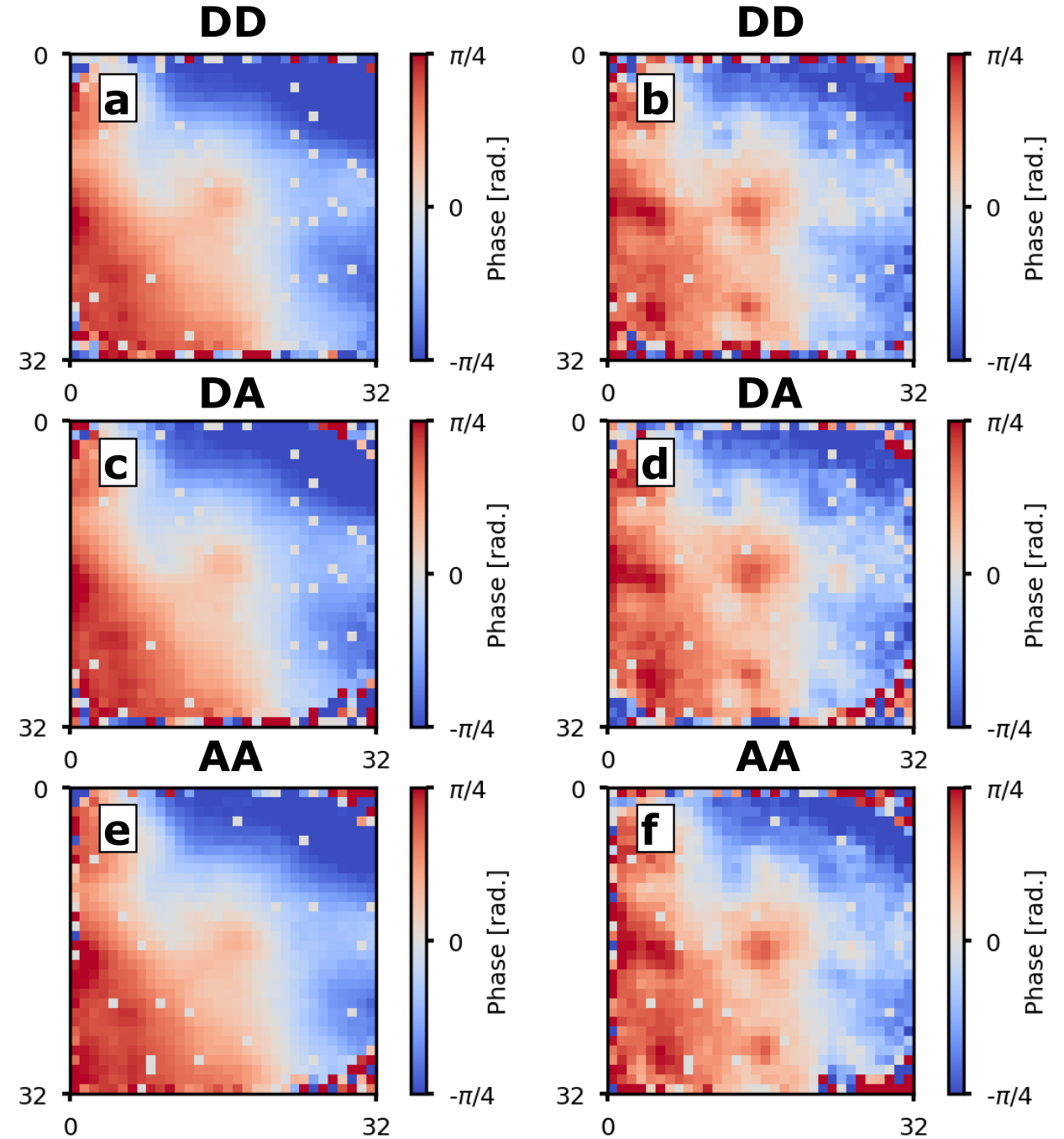}
    \caption{Retrieved phase images for non-birefringent test sample measurement. \textbf{(a)} and \textbf{(b)} Background and sample retrieved phase for $\bra{DD}$ polarization projection. \textbf{(c)} and \textbf{(d)} Background and sample retrieved phase for $\bra{DA}$ polarization projection. \textbf{(e)} and \textbf{(f)} Background and sample retrieved phase for $\bra{AA}$ polarization projection.}
    \label{fig:backgroundPhaseSubtr_NBP}
\end{figure}

While future developments will focus on using the SLM to physically correct this background phase, in this work we simply acquired a phase image of the background, as well as the sample and then numerically subtracted the background. For the measurement of the non-birefringent phase, the retrieved phase images of the background for the $\bra{DD}$, $\bra{DA}$, and $\bra{AA}$ polarization projections are shown in Fig.~\ref{fig:backgroundPhaseSubtr_BP}a, c, and e respectively. The respective retrieved phase images of the non-birefringent sample (before subtracting the background phase), again for polarization projections $\bra{DD}$, $\bra{DA}$, and $\bra{AA}$ are shown in Fig.~\ref{fig:backgroundPhaseSubtr_BP}b, d, and f. We first used Eq.~(\ref{eq:combinephi_NOON_SM}) to obtain a combined phase image of the background and the sample phase, and then subtracted the combined background phase image from the combined sample phase image. This yielded the results shown in Fig. 3c of the main text of this article. Likewise, Fig.~(\ref{fig:backgroundPhaseSubtr_NBP}) shows the background and sample retrieved phase images for the three polarization projections. We again used Eq.~(\ref{eq:combinephi_NOON_SM}) to combine the background phase images (Fig.~\ref{fig:backgroundPhaseSubtr_NBP}a, c, e) and the sample phase images (Fig.~\ref{fig:backgroundPhaseSubtr_NBP}b, d, f), and then subtracted the combined background phase image from the combined sample phase image. This yielded the results presented in Fig. 4c of the main text of this article.


\section{Image similarity between SLM test sample and retrieved phases}

\begin{figure}
    \centering
    \includegraphics[width=0.7\textwidth]{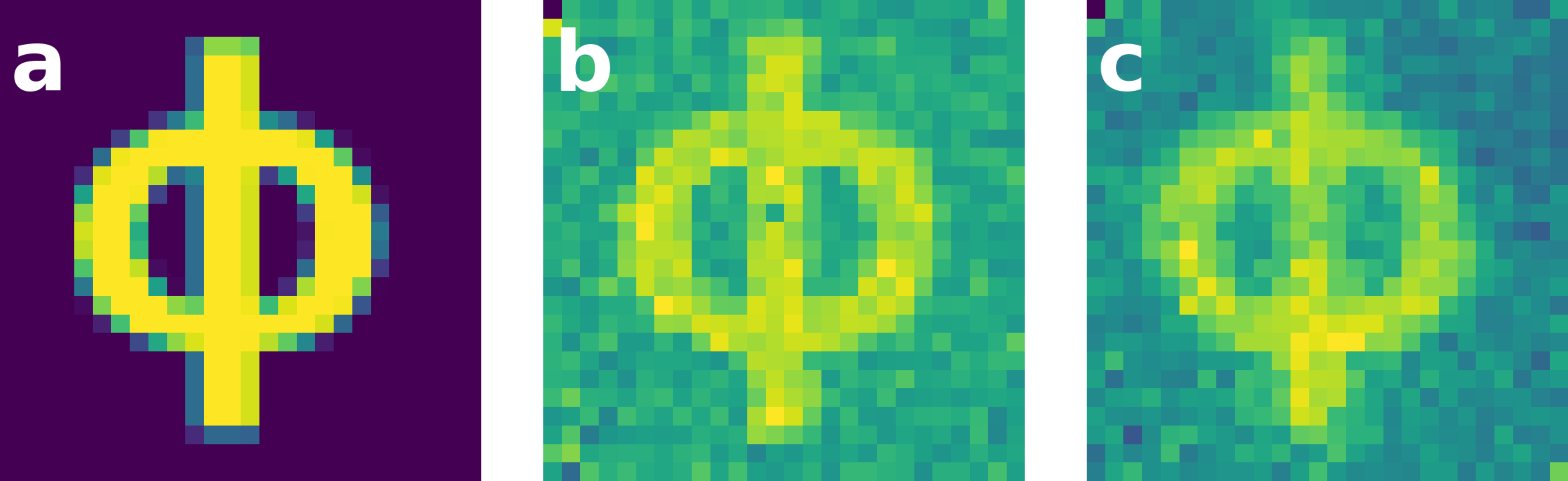}
    \caption{Images used for ZNCC calculation. \textbf{(a)} Central ROI for downsampled image of test sample applied to SLM. \textbf{(b)} Central ROI for phase image retrieved by classical measurement. \textbf{(c)} Central ROI for phase image retrieved by entanglement-enhanced measurement.}
    \label{fig:zncc_SLMphi}
\end{figure}

We calculated the image similarity between the birefringent test sample applied to the SLM, and the phase images retrieved by classical and entanglement-enhanced interferometric measurements (as shown in Fig. 3 of the main text of this article). The image similarity was quantified using the zero mean normalized cross-correlation (ZNCC), which is widely used in image template-matching ({\it 55\/}). The ZNCC parameter $ R_\mathrm{ZNCC} $ varies between -1 and 1. The higher $ R_\mathrm{ZNCC} $ is, the more similar two images $ I $ and $ J $ are, with
\begin{equation}
    R_\mathrm{ZNCC} = \frac{ \braket{ I(x,y) - \bar{I} | J(x,y) - \bar{J} } }{ \sqrt{\braket{ I(x,y) - \bar{I} | I(x,y) - \bar{I} } \braket{ J(x,y) - \bar{J} | J(x,y) - \bar{J} } } } \, .
\end{equation}
For the ZNCC calculations, we downsampled the image of the Greek letter ``$\phi$" applied to the SLM, such that it would have the same pixel resolution as the retrieved phase images measured with our SPAD array camera. We then calculated $ R_\mathrm{ZNCC} $ between the downsampled SLM image (Fig.~\ref{fig:zncc_SLMphi}a) and the retrieved phase image for the classical measurement (Fig.~\ref{fig:zncc_SLMphi}b) and for the entanglement-enhance measurement (Fig.~\ref{fig:zncc_SLMphi}c) respectively. Note that, as seen in Fig.~\ref{fig:zncc_SLMphi}b and c, for the retrieved phase images we used only a central region of interest (ROI) in order to avoid noisy edge effects. We obtained image similarity values $ R_\mathrm{ZNCC} $ between the image applied to the SLM and the classically retrieved phase image of 0.823, and between the image applied to the SLM and the entanglement-enhanced retrieved phase image of 0.848. This demonstrates that the classical and entanglement-enhanced measurements retrieved the sample phase approximately as well as each other.

\end{document}